\documentclass[11pt,preprint]{aastex}
\usepackage{times}
\usepackage{amssymb}
\usepackage{amsmath}
\usepackage{xspace}
\usepackage{natbib}

\newcommand{\degg}{\hbox{$^\circ$}\xspace}

\newcommand{\xmm}{\emph{XMM-Newton}\xspace}

\newcommand{\chandra}{\emph{Chandra}\xspace}

\newcommand{\Msun}{\hbox{$\rm\thinspace M_{\odot}$}}
\newcommand{\ls}
{\mathrel{\hbox{\rlap{\hbox{\lower4pt\hbox{$\sim$}}}\hbox{$<$}}}}
\newcommand{\gs}
{\mathrel{\hbox{\rlap{\hbox{\lower4pt\hbox{$\sim$}}}\hbox{$>$}}}}

\newcommand{\logxi}{\log (\xi/\rm{erg\,cm\,s}^{-1})}

\received{}
\begin{document}
\title{A deep X-ray view of the bare AGN Ark 120. I. Revealing the Soft X-ray Line Emission} 
\shorttitle{Soft X-ray emission in Ark 120}
\shortauthors{Reeves et al.}
\author{J. N. Reeves\altaffilmark{1,2}, D. Porquet\altaffilmark{3}, V. Braito\altaffilmark{1,4}, E. Nardini\altaffilmark{2}, A. Lobban\altaffilmark{5}, T. J. Turner\altaffilmark{6}}

\altaffiltext{1}{Center for Space Science and Technology, 
University of Maryland 
Baltimore County, 1000 Hilltop Circle, Baltimore, MD 21250, USA: jreeves@umbc.edu}
\altaffiltext{2}{Astrophysics Group, School of Physical and Geographical Sciences, Keele 
University, Keele, Staffordshire, ST5 5BG, UK; j.n.reeves@keele.ac.uk}
\altaffiltext{3}{Observatoire Astronomique de Strasbourg, Universit{\'e}
  de Strasbourg, CNRS, UMR 7550, 11 rue de l'Universit{\'e}, F-67000
  Strasbourg, France}
\altaffiltext{4}{INAF - Osservatorio Astronomico di Brera, Via Bianchi 46 I-23807 Merate (LC), Italy}
\altaffiltext{5}{Dept of Physics and Astronomy, University of Leicester, University Road, Leicester LE1 7RH, UK}
\altaffiltext{6}{Department of Physics, 
University of Maryland Baltimore County, 1000 Hilltop Circle, Baltimore, MD 21250, USA}

\begin{abstract}

The Seyfert 1 galaxy, Ark 120, is a prototype example of the so-called 
class of bare nucleus AGN, whereby there is no known evidence for the 
presence of ionized gas along the direct line of sight. Here deep ($>400$\,ks exposure), 
high resolution X-ray spectroscopy of Ark\,120 is presented, from {\it XMM-Newton} observations 
which were carried out in March 2014, together with simultaneous {\it Chandra}/HETG exposures. 
The high resolution spectra
confirmed the lack of intrinsic absorbing gas associated with Ark\,120, with the only X-ray absorption present 
originating from the ISM of our own Galaxy, with a possible slight enhancement 
of the Oxygen abundance required with respect to the expected ISM values in the Solar neighbourhood. 
However, the presence of several soft X-ray emission lines are revealed for the first time in the {\it XMM-Newton} 
RGS spectrum, associated 
to the AGN and 
arising from the He and H-like ions of N, O, Ne and Mg. The He-like line profiles of N, O and Ne appear velocity 
broadened, with typical FWHM widths of $\sim5000$\,km\,s$^{-1}$, whereas the H-like profiles are unresolved. 
From the clean measurement of the He-like triplets, we deduce that the broad lines arise from gas of density 
$n_{\rm e}\sim10^{11}$\,cm$^{-3}$, while the photoionization calculations infer that the emitting gas covers 
at least 10\% of $4\pi$ steradian.
Thus the broad soft X-ray profiles appear coincident with an X-ray component 
of the optical--UV Broad Line Region on sub-pc scales, whereas the narrow profiles originate on larger pc scales, 
perhaps coincident with the AGN Narrow Line Region. The observations show that Ark\,120 is not intrinsically 
bare and substantial X-ray emitting gas exists out of our direct line of sight towards this AGN.

\end{abstract}


\section{Introduction}

Photo-ionised or ``warm'' absorbers are commonly observed in at least 50\% of the 
UV/X-ray spectra of Seyfert 1s and type-1 QSO and are an important constituent of AGN 
(Reynolds 1997, Crenshaw, Kraemer \& George 2003, Porquet et al.\ 2004, Blustin et al. 2005, McKernan et al. 2007, Turner \& Miller 2009). 
Indeed the Seyfert warm absorbers that are frequently observed at 
high spectral resolution with {\sl XMM-Newton} and {\sl Chandra} are now known to give rise to 
numerous narrow absorption lines, usually blue-shifted -- implying outflowing winds -- 
of a few hundred km\,s$^{-1}$ up to a few thousand  km\,s$^{-1}$. These arise  
from various elements over a wide range of ionisation levels and column densities, especially 
from iron, oxygen, carbon, nitrogen, neon, silicon and sulfur (Kaastra et al. 2000, Kaspi et al., 2002, 
McKernan et al. 2003, Blustin et al. 2003). 
Signatures range from the lowly ionised  
Unresolved Transition Array (UTA) of M-shell iron ($<$ Fe\,\textsc{xvii})
at $\sim16-17$\AA~ (Sako et al. 2001, Behar et al. 2001), to absorption from highly ionised 
(H-like and He-like) iron which may originate from an accretion disk wind 
(Reeves et al. 2004, Risaliti et al. 2005a, Braito et al. 2007, Turner et al., 2008, Tombesi et al. 2010, 
Gofford et al. 2013). These spectroscopic measurements can
reveal crucial information on the outflow kinematics, physical conditions and locations 
relative to the central continuum source -- ranging from the inner nucleus (0.01 pc) to the 
galactic disk or halo (10 kpc).

However a small class of nearby Seyfert galaxies exist which show 
no (or very little) X-ray or UV absorption. These AGN are the
so-called {\it ``bare nucleus''} Seyferts or bare AGN. 
In principle the lack of intrinsic absorbing gas in these 
bare AGN allows a clean measurement of the innermost regions of the AGN and 
of the central engine 
closest to the black hole, removing any uncertainties as to how the absorbing 
gas is modeled.
Ark\,120 (or Arakelian\,120) is a nearby ($z=0.032713$, Osterbrock \& Phillips 1977, Theureau et al. 2005) and X-ray bright 
($F_{0.5-10\,{\rm keV}}=5.3\times10^{-11}$\,erg\,cm$^{-2}$\,s$^{-1}$, 
$F_{14-195\,{\rm keV}}=7.0\times10^{-11}$\,erg\,cm$^{-2}$\,s$^{-1}$; Patrick et al. 2011, Baumgartner et al. 2013) bare 
nucleus Seyfert 1. 
Along with its sister AGN, Fairall 9 (Emmanoulopoulos et al. 2011, Lohfink et al. 2012), 
it is the prototype example of a bare AGN.
Indeed it is one of the brightest and cleanest bare AGN known, displaying neither intrinsic reddening in
its IR/optical continuum nor evidence for absorption in UV and X-rays (Crenshaw et al.\ 1999, 
Reynolds 1997), allowing a clear view of the 
innermost regions of the AGN. A further key advantage for 
studying Ark\,120 is that it has a well determined reverberation based black 
hole mass, of $M_{\rm BH}$=1.5$\times$10$^{8}$\,M$_{\odot}$ 
(Peterson et al. 2004). 

An open question is whether Ark\,120 is intrinsically 
bare and devoid of circumnuclear X-ray emitting and/or absorbing gas,
 which may pose a challenge for unified schemes of AGN that imply the existence of 
wide scale obscuring and emitting gas (Antonucci 1993).
Indeed, Vaughan et al. (2004) presented 
an initial 100\,ks {\it XMM-Newton} observation of Ark\,120 in 2003, which from the spectra 
obtained with the RGS (Reflection Grating Spectrometer, den Herder et al. 2001) 
showed no significant soft X-ray emission or absorption features associated to the AGN. 
Furthermore the X-ray continuum was found to be smooth from the soft X-ray band up to 10 keV, 
with a large but featureless soft X-ray excess present at energies below 2\,keV. 
This was also confirmed in a {\it Suzaku} study by Nardini et al. (2011), who favored a 
relativistic accretion disk reflection origin for the soft X-ray excess and broad iron K$\alpha$ line. 
In an alternative explanation for the broad band spectrum, Tatum et al. (2012) 
accounted for the iron K$\alpha$ emission through Compton scattering off an accretion disk wind, 
which had to be viewed out of the direct line of sight in this AGN. 
Most recently, Matt et al. (2014) presented a simultaneous $\sim 100$\,ks 
{\it XMM-Newton} and {\it NuSTAR} observation of 
Ark\,120 obtained in 2013 and showed that the bare broad-band X-ray 
spectrum could be explained by Comptonization of UV photons through a warm scattering medium 
associated to an accretion disk corona.

This paper is the first of a series of papers to report upon
the analysis of an unprecedented deep observational campaign on Ark\,120, which was subsequently obtained
with {\it XMM-Newton} in 2014, with a total exposure exceeding 400\,ks (PI, D. Porquet). 
Part of the long {\it XMM-Newton} observations were performed simultaneously with {\it NuSTAR} to provide 
broad band hard X-ray coverage and with the High Energy Transmission Grating (HETG, Canizares et al. 2005) 
on-board {\it Chandra} to provide a high resolution view of the iron K band region. 
Here we concentrate on the high signal to noise and high resolution soft X-ray spectrum obtained with the 
RGS spectrometer on-board {\it XMM-Newton}. The primary goal is to determine 
whether the soft X-ray spectrum of Ark\,120 is intrinsically bare and devoid of circumnuclear X-ray gas, 
or indeed whether there are any signatures of ionized emission or absorption, 
which could arise from the accretion disk, the AGN broad and narrow line regions or from a nuclear outflow. 
We also present a search for any soft X-ray emission lines at high resolution 
from the Chandra/HETG above 1\,keV from Mg, Si, S.
Subsequent papers  will report in detail on the 
modeling of the iron K$\alpha$ line profile obtained as part of these observations (paper II, Nardini et al. 2016), as well 
as the nature of the broad band UV to hard X-ray continuum of Ark\,120 (paper III, Porquet et al. 2016, in preparation).
 
The paper is organized as follows. In Section\,2, we describe the analysis of the RGS observations, 
while in Section\,3 the overall properties of the 
soft X-ray RGS spectrum are presented. 
Section\,4 is devoted to the analysis of the Galactic ISM absorption towards Ark\,120 and 
the subsequent modeling of the soft X-ray continuum. Section\,5 then describes the first detection of soft X-ray 
emission lines from Ark\,120 that have now been made possible through the deep RGS exposure and Section\,6 
discusses their potential origin 
in the broad and narrow lined regions from the AGN. Values of H$_{\rm 0}=70$\,km\,s$^{-1}$\,Mpc$^{-1}$,
and $\Omega_{\Lambda_{\rm 0}}=0.73$ are assumed throughout and errors are quoted at 90\% 
confidence ($\Delta\chi^{2}=2.7$), for 1 parameter of interest. All spectral parameters
are quoted in the rest-frame of the AGN, at $z=0.032713$, unless otherwise stated. A conversion between energy and 
wavelength of $E = (12.3984$ \AA/$\lambda)$\,keV is adopted throughout.

\section{Observations and Data Reduction}\label{sec:obs}
\subsection{XMM-Newton Observations of Ark\,120}

XMM-Newton observed Ark\,120 four times between 18-24 March 2014, over 4 consecutive 
satellite orbits. Each observation was approximately 130\,ks in total duration, with the details 
of all four observations listed in Table\,1. First order dispersed spectra were obtained  
with the RGS and were reduced using the 
\textsc{rgsproc} script as part of the XMM-Newton SAS software v14.0. After screening the 
data for periods of high background, the net exposures for each RGS spectrum varied between 
$102.8 - 116.6$\,ks (see Table\,1). 
Prior to spectral analysis, channels due to bad pixels on the RGS CCDs were ignored as well as the two malfunctioning 
CCDs for RGS\,1 and RGS\,2 respectively. This yielded net background subtracted count rates of between 
$0.653\pm0.003$\,s$^{-1}$ -- $0.795\pm0.003$\,s$^{-1}$ for RGS\,1 and 
$0.746\pm0.003$\,s$^{-1}$ -- $0.909\pm0.003$\,s$^{-1}$ for RGS\,2, with the brightest observation being 
the first one in the series of four consecutive XMM-Newton orbits. In each case, the background count rate 
represented less than 5\% of the total count rate and thus the RGS observations were dominated by the AGN.

Spectra from the RGS\,1 and RGS\,2 were then also combined into a single RGS\,1+2 spectrum for each 
of the four XMM-Newton sequences, having checked that the resultant spectra were consistent with each 
other before and after combining the two RGS modules. Note that the spectral response files were subsequently 
averaged over the effective areas of the two RGS modules. In order to provide a very simplistic parameterization of the 
RGS spectra, we fitted the 0.35--2.0\,keV energy range (or 6.20\AA -- 35.4\AA) with a continuum comprising a 
powerlaw plus blackbody component, whereby the blackbody emission provides an initial zeroth order description 
of the known soft excess towards this AGN (Vaughan et al. 2004, Nardini et al. 2011, 
Matt et al. 2014). An initial Galactic absorption of hydrogen column density of
$N_{\rm H}=9.8\times10^{20}$\,cm$^{-2}$ (Kalberla et al. 2005) was adopted, modeled with the 
``Tuebingen--Boulder'' absorption model (hereafter {\sc tbabs} in {\sc xspec}) using 
the cross--sections and Solar ISM abundances of Wilms et al. (2000). 

Figure\,1 shows the combined RGS\,1+2 spectra 
for each of the 4 sequences, whereby the spectral parameters are all consistent within the margin of error; 
the average photon index was found to be $\Gamma=2.09\pm 0.04$, a blackbody temperature of $kT=132\pm4$\,eV 
and a neutral absorbing column in the local ($z=0$) Galactic frame of $N_{\rm H} = (8.3\pm0.3)\times10^{20}$\,cm$^{-2}$, 
slightly lower compared to the expected Galactic neutral H\,\textsc{i} column from 21\,cm measurements above. 
Nonetheless the fit is statistically 
poor, with a reduced chi-squared of $\chi^{2}/{\rm dof} = 3197.2/2308 = 1.385$ (where dof is the number of degrees of freedom in the fit), with strong residuals around the 
position of the neutral O\,\textsc{i} edge as well as systematic-like residuals throughout the spectrum. 
However it can be seen that all 4 sequences are consistent with each other, aside from some small differences 
in their absolute fluxes, with the brightest spectrum corresponding to the first \xmm\ observation (0.4--2.0\,keV band 
flux, $F_{\rm 0.4-2.0\,keV}=3.31\pm0.02 \times 10^{-11}$\,erg\,cm$^{-2}$\,s$^{-1}$) and the faintest 
corresponding to the 2nd and 4th observations as above ($F_{\rm 0.4-2.0\,keV}=2.72\pm0.02 \times 10^{-11}$\,erg\,cm$^{-2}$\,s$^{-1}$).

As the four spectra are consistent in shape, aside from a small difference in absolute flux of $\pm10$\% 
between the observations, they were subsequently combined to produce a single, deep, high signal to noise 
RGS spectrum of Ark\,120. 
This yielded net source count rates of $0.705\pm0.001$\,s$^{-1}$ and $0.808\pm0.001$\,s$^{-1}$ for each of the combined RGS\,1 and RGS\,2 spectra, with net exposures of 431.9\,ks and 
430.8\,ks respectively. 
Note the total number of source
counts over all four sequences, obtained from combining both of the RGS\,1+2 modules, is $>6.5\times10^{5}$ counts, 
with an effective total exposure of 862.7\,ks 
providing a very high S/N soft X-ray spectrum of Ark\,120. The subsequent combined spectrum was binned into 
$\Delta \lambda=0.03$\,\AA\ bins, which over-samples the RGS spectral resolution by at least a factor of 
$\sim \times 2$ compared to the FWHM resolution. Due to the high count rate statistics, $\chi^{2}$ 
minimization was employed in the subsequent spectral fitting, as typically there are $\sim700$\,counts per $0.03$\AA\ 
resolution bin, corresponding to a S/N of $>25$ per bin. 

\subsection{Chandra HETG Observations of Ark\,120}

The High Energy Transmission Grating (HETG) onboard \chandra\ 
also observed Ark\,120 from 17 March to 22 March 2014. 
Due to scheduling constraints, the \chandra\ observations were split into 3 sequences, 
overlapping with the 1st, 2nd and 3rd of the \xmm\ sequences, with the 2nd sequence 
shorter than the other two, see Table\,1 for details.
Spectra were extracted with the \textsc{ciao} package v4.3. 
Only the first order dispersed spectra were considered for both the MEG (Medium Energy Grating) and 
HEG (High Energy Grating) and the $\pm1$ orders for each grating were subsequently combined 
for each sequence. No significant spectral variability was observed between the 3 sequences and the 
spectra were consistent, with only modest $\sim 10$\% variations in source flux. 
Therefore the spectra were combined from all three sequences to yield a single 1st order spectrum 
for each of the MEG and HEG, yielding respective net source count rates of $0.868\pm0.003$\,s$^{-1}$ 
and $0.491\pm0.002$\,s$^{-1}$ respectively for a total exposure time of 120.5\,ks. 
Thus the total counts obtained exceeded $1.0\times10^{5}$ and $5\times10^{4}$ counts
for MEG and HEG respectively. Note that the background contribution towards the count rate was negligible. 
Due to the high flux ($7\times10^{-11}$\,erg\,cm$^{-2}$\,s$^{-1}$ from 0.5-10\,keV) and high count rates obtained 
from Ark\,120, the zeroth order image and spectra were not usable due to severe pile-up of the central source.

The resultant 2014 1st order
source spectra were subsequently binned to $\Delta \lambda = 0.01$\,\AA\ and 
$\Delta \lambda = 0.005$\,\AA\ bins for 
MEG and HEG respectively, which over-samples their respective FWHM spectral resolutions by a factor of $\times2$. 
The C-statistic was 
employed in the subsequent spectral fits with the HETG, as although the overall count rate is high, towards 
the lower energy (longer wavelength) 
end of each grating spectrum the total source counts per bin drops below $N<20$ 
in some bins. In the case of $\chi^{2}$ minimization, this would lead to the continuum level being 
somewhat underestimated at soft X-ray energies.

\section{The Soft X-ray Spectrum of Ark\,120}\label{sec:initial}

Initially we concentrated on the analysis of the time-averaged 2014 \xmm\ RGS observations. 
Figure\,2 shows the overall 
2014 $\nu F_{\nu}$ RGS spectrum of Ark\,120, fluxed against a powerlaw of $\Gamma=2$ in 
the soft X-ray band. 
The spectrum shown in Figure 2 is largely devoid of any strong ionized absorption lines in the AGN rest frame, 
as might be expected for Ark\,120 given its past record as a bare Seyfert 1 galaxy and the lack of 
any warm absorber in the AGN (e.g. Vaughan et al. 2004).
The expected rest frame positions of several $1s-2p$ resonance lines 
(e.g from the He and H-like ions of N, O, Ne and Mg) are marked in the figure and there appears to be no significant 
absorption lines at any of these positions. However in the local ($z=0$) observed frame, there appears to be several 
absorption features, in particular around the position of the neutral O\,\textsc{i} edge at $\sim 23$\,\AA\ there is a strong resonance absorption line (at 23.5\AA\ or 527\,eV) which may be identified with the K$\alpha$ absorption line 
due to O\,\textsc{i} in our galaxy (Gorczyca et al. 2013). In addition another strong absorption line is observed 
near 31.2\AA, which likewise may be attributed to the neutral K$\alpha$ resonance line of N. Thus it appears clear 
we need to account for the absorption in the ISM due to our own galaxy in the soft X-ray spectrum of Ark\,120, 
which we investigate below. Furthermore, although there appears little in the way of intrinsic ionized absorption
towards Ark\,120, there may be some indication ionized X-ray {\it emission} lines associated with the AGN, 
for instance associated with O\,\textsc{vii}, O\,\textsc{viii}, Ne\,\textsc{ix} or Mg\,\textsc{xii}. 

\section{X-ray Absorption in the Galactic ISM towards Ark\,120}

Before investigating the presence of any ionized gas in emission (or absorption) intrinsic to Ark\,120, we first attempt to model the absorption associated with the line of sight through the ISM of our own Galaxy. Due to the relatively 
low Galactic latitude of Ark\,120 ($b=-21\degg.1$), the Galactic H\,\textsc{i} column is thought to be relatively 
high, with $N_{\rm H} = 9.78\times10^{20}$\,cm$^{-2}$ as measured from 21\,cm surveys (Stark et al. 1992, Kalberla
et al. 2005) and thus this absorption needs to be modeled before we can accurately determine the intrinsic 
emission properties of the AGN. The Galactic hydrogen column was allowed to vary, noting that there can also 
be an additional contribution associated with molecular hydrogen (e.g. Willingale et al. 2013)

We fitted the combined mean 2014 RGS spectrum, adopting a powerlaw plus blackbody form to 
parameterize the soft X-ray continuum of Ark\,120. As the RGS bandpass contains a contribution from both the hard X-ray 
powerlaw extending up to higher energies up to at least 70\,keV (e.g. Matt et al. 2014, Porquet et al. 2016), as well as the prominent soft excess below 2\,keV, we adopt both components in order to account for any spectral curvature present in the soft X-ray band.
The AGN continuum emission was parameterized 
by a photon index of $\Gamma=2.15\pm0.05$ and a blackbody temperature of $kT=110\pm10$\,eV.
For the Galactic absorption, we initially adopted the \textsc{tbvarabs} model of Wilms et al. (2000), which accounts 
for the photoelectric absorption edges due to abundant elements in the X-ray band and allows for the possibility of 
variable abundances compared to chosen tabulated Solar values. A neutral hydrogen column of 
$N_{\rm H}=(9.4\pm0.4) \times10^{20}$\,cm$^{-2}$ was found, close to the reported 21\,cm value above, 
with a relative O abundance compared to those in Wilms et al. (2000) or Asplund et al. (2009) 
of $A_{0} = 1.39\pm0.10$. 

However while this model provides a reasonable first order parameterization of the soft X-ray continuum 
of Ark\,120, the fit is statistically very poor, with a reduced $\chi^2$ corresponding to 
$\chi_{\nu}^{2} = \chi^{2}/{\rm dof} = 1561.3/938$. 
The \textsc{tbvarabs} or \textsc{tbabs} models only include the bound--free photoelectric absorption associated 
with the K and L-shell edges of abundant elements, but do not include any resonance absorption line structure due 
to the neutral ionization states of these elements. For instance Figure\,3 (upper panel) shows the fluxed RGS spectrum 
(obtained against a simple 
unabsorbed $\Gamma=2$ powerlaw model) around the 
neutral O edge region, with the above \textsc{tbvarabs} photoelectric absorption model super-imposed upon the spectrum. It is apparent that the
prominent O\,\textsc{i} K$\alpha$ absorption line observed near 23.5\,\AA\ (or 527\,eV) is left unmodelled, whilst in reality the drop in the 
spectrum around the O edge (observed between 22.7\AA -- 23.0\AA) is 
not sharp, which is likely due to the presence of several resonances around the edge threshold energy (see 
de Vries et al. 2003, Gorczyca et al. 2013, Gatuzz et al. 2014). 
Indeed if we parameterize the 23.5\,\AA\ absorption with a simple narrow ($\sigma=1$\,eV) Gaussian absorption line we 
obtain an observed frame line energy of $527.2\pm0.2$\,eV (or $\lambda = 23.52\pm0.01$\,\AA), which is in very good agreement with the expected energy of 
527.4\,eV, as also found from measurements towards Galactic binary systems (Gatuzz et al. 2014). 
Furthermore a second strong absorption line is present at $E=396.6\pm0.5$\,eV or $31.26\pm0.04$\,\AA\ 
(e.g. see Figure 2, lowest panel), which is in agreement with 
the expected position of the N\,\textsc{i} K$\alpha$ line at 396.1\,eV (Kaastra et al. 2011). The parameters of these Galactic 
absorption lines are also listed in Table\,2. Upon addition of both of these ad-hoc Galactic 
absorption lines, the fit statistic to the RGS data improves to $\chi_{\nu}^{2} = 1392.7/934$, 
although the model is still formally rejected by the data.

\subsection{A Comparison between ISM Absorption Models}

A more physical model was then adopted to model the Galactic ISM absorption towards Ark\,120, using the \textsc{tbnew} model, which is an  
improved higher resolution version of the original \textsc{tbabs} or \textsc{tbvarabs} ISM absorption model of Wilms et al. (2000). 
The \textsc{tbnew} model includes the 
various resonance absorption lines around the positions of the neutral O K-shell, Ne K-shell and 
Fe L-shell edges\footnote{see http://pulsar.sternwarte.uni-erlangen.de/wilms/research/tbabs/ for further details of the \textsc{tbnew} model.}.
The ISM abundance table of Wilms et al. (2000) was also adopted for the modeling. The power-law plus blackbody continuum form was retained 
from before, although the parameters are allowed to vary freely.
The goodness of fit obtained is subsequently improved compared to the simpler
\textsc{tbvarabs} model ($\chi_{\nu}^{2} = 1344.0/936$) and the model is able to self consistently 
account for the strong O\,\textsc{i} ${\rm K}\alpha$ absorption
line present observed at 527.2\,eV (or $\lambda = 23.52$\AA) in the $z=0$ frame. 
Figure 3 (lower panel) also shows a zoom-in around the O K-shell region, but with the best-fit \textsc{tbnew} 
overlaid on the fluxed RGS spectrum. In addition to the O\,\textsc{i} ${\rm K}\alpha$ absorption line, the model provides 
a better description of the data around the O K-shell edge, due to the various higher order resonance line structures, which produces a more gradual 
decrease at the position of the O K-shell edge. Figure 4 displays the same model, but now folded through the RGS instrumental response, versus 
the count rate spectrum and it can be seen that the model reproduces both the O\,\textsc{i} K$\alpha$ edge region.

We note here that the excess emission blue-wards of the O K-shell edge region is likely associated with emission from the O\,\textsc{vii} triplet, 
in the region from 561\,eV to 574\,eV in the AGN rest frame, as can be seen in Figure 4. Indeed several ionized emission lines are observed from the AGN in the RGS spectrum, 
predominantly from the He and H like ions for N, O, Ne and Mg and the fit statistic improves considerably upon their inclusion in the model to 
$\chi_{\nu}^{2} = 1087.3/914$. The soft X-ray emission line properties of Ark\,120 are discussed in detail in Section 5.

In addition the \textsc{tbnew} model also self consistently models the absorption around the neutral Fe L and Ne K edges respectively, for instance 
some structure around the Fe L edge is also visible at 17.5\AA\ (but much weaker than at Oxygen) in the fluxed spectrum in Figure 2. 
However the absorption line structure is not included around the N\,\textsc{i} K-shell edge region in \textsc{tbnew} 
and therefore we retain the simple Gaussian absorption line to parameterize the visible N\,\textsc{i} K$\alpha$ absorption line at 396.6\,eV.
We subsequently adopt the \textsc{tbnew} model of the Galactic absorption as our initial baseline model for the ISM absorption towards Ark\,120, with the baseline 
absorption and continuum (powerlaw plus blackbody) 
parameters listed in Table\,2. Note that while the O\,\textsc{i} K$\alpha$ absorption line is self consistently fitted 
in the \textsc{tbnew} model, the simple Gaussian line parameterization obtained from before is listed also in Table\,2 for completeness.

Overall the hydrogen column density obtained by the \textsc{tbnew} model is $N_{\rm H} = (8.8\pm0.4)\times10^{20}$\,cm$^{-2}$, which is slightly below the reported 21\,cm value of 
$N_{\rm H} = 9.78\times10^{20}$\,cm$^{-2}$. 
However the neutral Oxygen abundance relative to the ISM values collated in Wilms et al. (2000) or Asplund et al. (2009) 
(of ${\rm O}/{\rm H} = 4.90\times10^{-4}$) 
is left free to vary (other elements are fixed at Solar). With respect to this value, the O abundance along the line of sight in the 
Galactic ISM is found to be mildly super-Solar, with $A_{\rm O}=1.62\pm0.10$. If instead we adopt the earlier photospheric 
Solar abundance table of Grevesse \& Sauval (1998), 
which has a higher absolute oxygen abundance (of ${\rm O}/{\rm H} = 6.76\times10^{-4}$), then the relative O abundance compared to this value is lower 
($A_{\rm O}=1.17\pm0.08$), while the hydrogen column is consistent with the previous value ($N_{\rm H} = 9.0\pm0.4 \times10^{20}$\,cm$^{-2}$). 

We then compared the Ark\,120 soft X-ray spectrum with the \textsc{ismabs} absorption model (Gatuzz et al. 2014). In addition to the updated resonance line cross-sections around the O, Ne K-shell and Fe L-shell complexes, 
the model also can include the contribution from ISM absorption 
from once and twice ionized ions from most of the cosmically abundant elements, as has been observed towards both Galactic sources 
and AGN (Pinto et al. 2010, 2012). A further advantage of the \textsc{ismabs} model is that it allows the user to directly determine 
the column density of both the neutral atoms of abundant elements (as well as the ionized species), which can also be inferred indirectly 
from the \textsc{tbnew} model from the hydrogen column and then relative abundance ratio.

Applying the \textsc{ismabs} absorption model to the Ark\,120 RGS spectrum results in a very similar fit as per the \textsc{tbnew} model (with $\chi_{\nu}^{2} = 1086.6/911$)
and the comparison of the parameters between these two ISM models are tabulated in Table\,3. The hydrogen column density along the line of sight through our galaxy is also similar, with $N_{\rm H} = (8.5\pm0.4)\times10^{20}$\,cm$^{-2}$.
The neutral O column 
density was also allowed to vary, which primarily contributes towards the O\,\textsc{i} K$\alpha$ absorption line as well 
as the higher order absorption, as per the \textsc{tbnew} model. The neutral O column density is found to be 
$N_{\rm O} = (6.5\pm0.3)\times10^{17}$\,cm$^{-2}$, which thus corresponds to an abundance ratio of 
${\rm O}/{\rm H} = (7.7\pm0.5) \times10^{-4}$, which is a factor of $A_{\rm O}=1.57\pm0.10$ above the relative O/H abundance value
tabulated in Wilms et al. (2000), although again it is consistent with the higher O abundances tabulated by 
Anders \& Grevesse (1989) and Grevesse \& Sauval (1998).

The column densities of neutral N, Ne and Fe  were also allowed to vary, the best fit values obtained from the 
\textsc{ismabs} model are:- $N_{\rm N}=(7.2\pm2.7) \times 10^{16}$\,cm$^{-2}$, 
$N_{\rm Ne}=(1.2\pm0.3) \times 10^{17}$\,cm$^{-2}$ and $N_{\rm Fe}=(2.6\pm0.6) \times 10^{16}$\,cm$^{-2}$ respectively.
(see Table\,3). Note the ${\rm N}/{\rm H}$ and ${\rm Fe}/{\rm H}$ abundances are consistent with the Solar values 
in Wilms et al. (2000), as tabulated in Table 3, while the ${\rm Ne}/{\rm H}$ abundance appears slightly higher 
compared to these tabulated values.
All other elemental column densities were fixed according to their default Solar abundances in the \textsc{ismabs} model, which otherwise assumes values calculated from Grevesse \& Sauval (1998)\footnote{Note that \textsc{ismabs} 
also assumes that the He abundance is 10\% of that of hydrogen}, for a given hydrogen column. 

Finally there appears to be no evidence for any ionized absorption associated with the Galactic ISM towards Ark\,120. 
If for the \textsc{ismabs} model we allow the column densities of once ionized O (denoted O$^{+}$) and twice ionized O (denoted O$^{2+}$) 
to vary (at $z=0$), then only upper limits are found corresponding to $N_{\rm O^{+}}<4.2\times10^{16}$\,cm$^{-2}$ 
and $N_{\rm O^{2+}}<0.6\times10^{16}$\,cm$^{-2}$. These limits are at least a factor of $\times10$ and $\times100$ 
smaller respectively than the neutral O column through our Galaxy. Neither is there any evidence for any ISM absorption associated 
with the host galaxy of Ark\,120, in the rest frame at $z=0.032713$. The limit on the column of any neutral O 
absorption at $z=0.032713$ is $N_{\rm O}<2.5\times10^{16}$\,cm$^{-2}$, which translates into a hydrogen column 
of $N_{\rm H}<5\times10^{19}$\,cm$^{-2}$ assuming a Solar abundance of ${\rm O}/{\rm H}=4.9\times10^{-4}$ 
according to Wilms et al. (2000). This is perhaps not surprising upon inspection of the RGS data; at a redshift of 
$z=0.032713$, the O\,\textsc{i} K$\alpha$ absorption line at 527.2\,eV (or 23.52\AA) corresponds to an observed 
energy of 510.5\,eV (or 24.29\AA) and there is subsequently no evidence for absorption at these positions in either 
Figure 3 or 4.

\subsection{Dependence on the Soft X-ray Continuum}

The above analysis suggests that at least the Oxygen abundance (and possibly Ne) in the Galactic ISM 
towards Ark\,120 could be 
somewhat higher than some of the tabulated Solar ISM values (Wilms et al. 2000, Lodders 2003, Asplund et al. 2009). 
However while the column densities of O, Ne and Fe are well determined from the discrete atomic 
features (absorption lines and edges) in the 
RGS spectrum, the absolute hydrogen column density is inferred from the downwards curvature of the spectrum towards 
the lowest energies. Thus the measured $N_{\rm H}$ values may depend on how the soft X-ray continuum is modeled.
Therefore instead of the simple powerlaw plus blackbody continuum form, a continuum consisting of 
a powerlaw (responsible for the continuum above 2\,keV) and a Comptonized disk blackbody spectrum was adopted, 
parameterized 
in this case by the \textsc{comptt} model (Sunyaev \& Titarchuk 1985, Titarchuk 1994). The latter component represents 
the Compton upscattered Wien tail of the optically-thick inner disk emission, which may be produced by a warm scattering layer above the disk and it has been suggested that such an emission component could reproduce the mainly featureless 
UV to soft X-ray excesses observed in many AGN (Done et al. 2012, Jin et al. 2012). The main effect on the spectrum is that this produces a broader soft excess than a single unmodified blackbody component, as the latter would peak only 
in the UV band. 

Upon adopting instead \textsc{comptt} form for the soft X-ray excess, it is found that the hydrogen column density 
was somewhat higher compared to the powerlaw plus blackbody case. 
The absorption parameters are tabulated in Table\,3 and applied 
to both the \textsc{tbnew} and \textsc{ismabs} absorption models (denoted as \textsc{tbnew2} and \textsc{ismabs2} 
respectively in Table\,3). It can be seen that the neutral hydrogen column has increased for both cases, e.g. for the 
\textsc{tbnew} case, then $N_{\rm H} = (11.1\pm0.4)\times10^{20}$\,cm$^{-2}$, slightly higher than the Galactic value 
reported from 21\,cm measurements. On the other hand, the column density of neutral oxygen remained consistent 
regardless of which form is adopted for the soft X-ray continuum, i.e. for the \textsc{tbnew} absorption model, then 
$N_{\rm O} = (7.0\pm0.5)\times10^{17}$\,cm$^{-2}$ (for the blackbody case) vs. $N_{\rm O} = (6.5\pm0.3)\times10^{17}$\,cm$^{-2}$ (for the \textsc{comptt} case). Thus when the soft X-ray spectrum is modeled with 
the Comptonized disk spectrum, due to the higher implied $N_{\rm H}$ values, the relative abundances are lower, 
largely consistent with the Solar values. Thus for Oxygen, the relative abundances with respect to the Solar values of 
Wilms et al. (2000) are $A_{\rm O}=1.20\pm0.04$ (for \textsc{tbnew}) and $A_{\rm O}=1.10\pm0.08$ (for \textsc{ismabs})
respectively. 

\subsubsection{The Form of the Soft X-ray Excess}

Finally we note the continuum parameters associated with the \textsc{comptt} model. As the hard powerlaw constrained primarily above 2\,keV is not well determined, its photon index has been fixed $\Gamma=1.8$ (close to the best-fit 
value obtained for the EPIC-pn data, Nardini et al. 2016), 
with a corresponding photon flux (normalization) at 1\,keV 
of $N_{\rm PL} = (9.33\pm0.13) \times 10^{-3}$\,photons\,cm$^{-2}$\,s$^{-1}$. For the Comptonized disk 
spectrum, an input temperature for the seed photons of $kT_{\rm in} = 20$\,eV is assumed, which may be reasonable 
for the emission from the inner optically thick disk around a black hole of mass $1.5\times10^{8}$\Msun\ in Ark\,120, with a bolometric luminosity of $2\times10^{45}$\,erg\,s$^{-1}$, as determined from the broad-band SED (Porquet et al. 2016), 
which suggests accretion occurs at a rate of at least 10\% of the Eddington value.
These input seed disk photons are then assumed to be upscattered by a warm scattering medium, which we initially 
assume has an electron temperature of $kT_{\rm e}=2$\,keV; for this temperature the optical depth required to account 
for the soft excess is $\tau=3.6\pm0.3$. However $kT_{\rm e}$ and $\tau$ are largely degenerate upon each other; if instead 
we adopt a lower temperature of $kT_{\rm e}=0.5$\,keV for the Comptonizing electrons, then a higher optical depth of 
$\tau=8.5\pm0.5$ is obtained. In either case, the soft excess corresponds to between $\sim40-50$\% of the total unabsorbed 0.4--2.0\,keV band flux of $4.5\times10^{-11}$\,erg\,cm$^{-2}$\,s$^{-1}$. Note that the \textsc{comptt} model produces 
a similar goodness of fit compared to the simple powerlaw plus blackbody model, given the relatively narrow band of the RGS 
spectrum.

A further more detailed discussion into the nature of the soft excess
will be presented in a subsequent paper (Porquet et al. 2016), which will investigate the broadband 
XMM-Newton and {\it NuSTAR} observations, as well as the modeling of the optical to hard X-ray SED obtained 
from simultaneous data as part of the 2014 campaign. 

\section{The Emission Line Spectrum}\label{sec:emission}

Having adequately described the Galactic line of sight absorption towards Ark\,120, the intrinsic emission spectrum 
of Ark\,120 was then investigated. It is known that Ark\,120 is a bare Seyfert 1, which contains no intrinsic absorption 
associated with the AGN, such as a warm absorber, which is also confirmed later in Section\,5.5 from the contemporary 2014 
observations. However one open question is whether the AGN is truly bare, or indeed whether there is evidence for emission from photoionized gas surrounding the AGN instead of absorption along our line of sight through the AGN, as might 
be expected in the context of AGN orientation dependent Unified Schemes (Antonucci 1993, Urry \& Padovani 1995).

As a baseline model, the powerlaw plus blackbody continuum model was adopted, using the \textsc{tbnew} model for 
the Galactic ISM absorption, with parameters reported  in Table 2 and described in the previous section. Note that any dependency of the emission line spectrum on either the continuum form or the Galactic ISM model was checked and the 
subsequent parameters found to be consistent within errors. To provide an initial inspection of the contribution of any emission component, the summed 2014 RGS spectrum was first binned to a more coarse resolution of $\Delta\lambda=0.1$\AA\ per spectral bin. This is a little worse than the intrinsic resolution of the RGS, which is typically 
$\Delta\lambda=0.06-0.08$\AA\ (FWHM) over the 6--35\AA\ bandpass, but serves to highlight any prominent emission lines in the spectrum.

The RGS spectrum binned to this resolution is shown in Figure 5. The baseline continuum and Galactic 
absorption model overall provides a very good description of the shape of the RGS spectrum (see upper panel), however several strong positive deviations at $>3\sigma$ level are apparent in the lower panel. 
Indeed the fit statistic of the baseline model to the binned RGS spectrum is poor, 
with $\chi_{\nu}^{2} = 464.4/273 = 1.701$, 
formally rejected with a corresponding null hypothesis 
probability of $4.1\times10^{-12}$. Indeed the addition of seven Gaussian emission lines to the model 
significantly improves the fit statistic to $\chi_{\nu}^{2} = 296.3/254 = 1.165$, which is now statistically 
acceptable (null hypothesis probability of $3.5\times10^{-2}$). The centroids of the Gaussian profiles are close to several strong expected K-shell lines of N through to Mg; their centroid energies and likely identifications are:-  $502\pm2$\,eV (N\,\textsc{vii} Ly$\alpha$), $568\pm2$\,eV (O\,\textsc{vii} He$\alpha$), $654\pm1$\,eV (O\,\textsc{viii} Ly$\alpha$), $865\pm5$\,eV (Fe\,\textsc{xviii} $3d\rightarrow 2p$ or 
O\,\textsc{viii} RRC), $905\pm4$\,eV (Ne\,\textsc{ix} He$\alpha$), $1343\pm9$\,eV (Mg\,\textsc{xi} He$\alpha$) and 
$1476\pm9$\,eV (Mg\,\textsc{xii} Ly$\alpha$). Thus most of the emission lines appear to be associated with the 
He and H-like transitions of abundant elements. Several of the He-like lines appear to be resolved, especially 
for O\,\textsc{vii}, which when modeled with a single Gaussian requires a line width of $\sigma=6.8\pm1.4$\,eV. 
If these line widths are intrinsic, they would imply a significant degree of velocity broadening, with FWHM 
values in the range $\sim 5000-8500$\,km\,s$^{-1}$. On the other hand, the widths may be the result of a blend 
of contributions from the forbidden, intercombination and resonance emission components associated with 
the triplet emission, which is investigated further in Section\,5.3.

\subsection{The High Resolution Line Spectrum}

In order to provide a more accurate parameterization of the emission line spectrum and the line profiles, we 
reverted back to the spectrum binned at a finer resolution which adopted spectral bins 
of width $\Delta\lambda=0.03$\,\AA\, over-sampling the FWHM resolution of the RGS by a factor of $\times 2-3$. 
The multiple spectral panels in Figure\,6 show the data/model ratio of the RGS spectrum at this resolution to the baseline absorbed ({\sc tbnew}) continuum model, over the 0.4--1.6\,keV 
range, whereby the expected rest energies or wavelengths of the main emission lines are marked with vertical dashed lines 
on the panels.  The above emission line profiles are revealed in the residuals, in particular spectral structure is resolved around the He-like triplets especially from O\,\textsc{vii} and Ne\,\textsc{ix}. 
Furthermore positive residuals are also present in the region of the He-like N\,\textsc{vi} triplet around $\sim430$\,eV, 
although this part of the spectrum has been more coarsely binned by a factor of $\times 3$ as the signal to noise of the 
spectrum declines below 0.5\,keV. 

As before, the fit statistic improves considerably, now 
upon the addition of 8 Gaussian line profiles including the emission from N\,\textsc{vi}, with the fit statistic 
decreasing substantially from $\chi_{\nu}^{2} = 1344.0/936$ without any line emission (null hypothesis probability 
of $3.6\times10^{-17}$) to $\chi_{\nu}^{2} = 1087.3/914$ with the lines included. 
Initially the He-like line profiles have been modeled with single broad Gaussian emission lines and 
their properties, along with the H-like emission lines, are reported in Table\,4. For instance the O\,\textsc{vii} profile 
is resolved with a width of $\sigma=6.8^{+1.5}_{-1.2}$\,eV, noting that the fit is significantly worse (by $\Delta \chi^2 = 68.1$) if the line width is restricted to $\sigma<1$\,eV. Similar widths are also derived from the other 
He-like profiles (although Mg\,\textsc{xi} is less well determined, as is its centroid energy). The centroid energy of the O\,\textsc{vii} line at $568.2^{+1.7}_{-1.6}$\,eV suggests an identification with the intercombination emission from He-like Oxygen, which may imply a high electron density (Porquet \& Dubau 2000, Porquet et al. 2010), 
which is discussed further later. 
The N\,\textsc{vi} triplet (seen in the upper panel of Figure 6) 
is similar in this regard, the line centroid energy of $428.4\pm2.3$\,eV is consistent with the majority of the emission originating from the intercombination transitions (near 426.3\,eV), but is inconsistent with the expected energies of the forbidden 
(at 419.8\,eV) and resonance (430.7\,eV) emission lines. Conversely the centroid energy of the He-like triplet of 
Ne\,\textsc{ix} measured at $904.8\pm3.6$\,eV is consistent with the expected energy of the forbidden line at 905.1\,eV, 
but appears inconsistent with the positions of the intercombination and resonance lines at 914.8\,eV and 
922.0\,eV respectively. The modeling of the He-like triplets and the question of whether they are intrinsically velocity broadened or instead arise from a combination of the forbidden, intercombination and resonance components is discussed later in Section 5.3.

The H-like lines from N\,\textsc{vii}, O\,\textsc{viii} and Mg\,\textsc{xii} are all significantly detected close to their expected rest frame energies or wavelengths, with only the H-like line 
of Ne\,\textsc{x} not being formally detected, as listed in Table\,4. 
At first sight, unlike the emission from the He-like ions, 
the H-like lines generally appear to be unresolved, with the corresponding upper-limits given to their $1\sigma$ Gaussian widths in Table 4. The H-like profiles 
are also discussed in further detail in the next section. The only uncertain identification is for the emission measured at 
$864.7\pm1.9$\,eV, which may be associated to L-shell emission from iron (e.g. Fe\,\textsc{xviii} $3p \rightarrow 2s$), 
but could also contain a contribution from the radiative recombination continuum (RRC) of O\,\textsc{viii}.

None of the higher order lines are detected, generally with tight upper limits. For instance 
the upper-limit on the flux from the O\,\textsc{vii} He$\beta$ emission ($E=661-666$\,eV) 
is $<0.92\times10^{-5}$\,photons\,cm$^{-2}$\,s$^{-1}$ (or an equivalent width of $<0.2$\,eV), which 
implies a ratio of the ${\rm He}\alpha/{\rm He}\beta$ emission of $>20$. Likewise the ratio of fluxes of O\,\textsc{viii} 
Ly$\alpha$ to Ly$\beta$ lines is $>5$. This may imply the line emission is associated with recombination (and subsequent radiative cascade) following photoionization rather than radiative decay following photo-excitation, as the 
latter process generally boosts the strengths of the higher order (e.g. Ly$\beta$ and He$\beta$) emission lines (see Kinkhabwala et al. 2002). If this is the case, detectable RRC emission may be expected. The data are consistent 
with the presence of the O\,\textsc{viii} RRC as noted above, with a total flux of 
$1.4\times10^{-5}$\,photons\,cm$^{-2}$\,s$^{-1}$, 
while the limit obtained on the temperature is $kT<10$\,eV (or $<6\times10^{4}$\,K). Formally the 
O\,\textsc{vii} RRC is not detected, with an upper limit of $<2.3\times10^{-5}$\,photons\,cm$^{-2}$\,s$^{-1}$ 
on the photon flux. Note that in the latter case, it may be difficult to detect an RRC feature, due to the 
large velocity width of the O\,\textsc{vii} emission, resulting in a weak and broadened profile. Overall, the 
line emission is consistent with a photoionized plasma, a view supported by the \textsc{xstar} emission 
modeling presented later in Section\,6.4.

The X-ray spectrum does not allow for any strong contribution from a collisionally ionized plasma, which appears  consistent with the lack of any strong iron L-shell emission lines. Furthermore, any resonance components in the He-like 
triplets are weak (Section\,5.3), 
which generally implies a low temperature plasma. For a $kT=0.5$\,keV temperature plasma with 
Solar abundances, the upper limit on the luminosity of any emission from a collisionally ionized plasma is 
$<2.5\times10^{41}$\,erg\,s$^{-1}$, which is $<0.25$\% of the total 0.4--2.0\,keV band soft X-ray luminosity from the AGN. 
In comparison the total luminosity derived from the Gaussian emission line profiles is $\sim1.0\times10^{42}$\,erg\,s$^{-1}$, 
thus any thermal plasma contributes less than 25\% of the total line emission. In practice, the contribution of a collisionally ionized plasma towards 
the observed emission lines is likely much smaller, as a significant proportion of the luminosity of any collisional plasma arises 
in the thermal bremsstrahlung continuum, rather than in the lines themselves.


\subsection{Emission Line Profiles}

In order to derive further constraints on the kinematics of the line emitting gas, velocity profiles of six 
of the strongest emission lines were constructed, from adopting the data/continuum model ratios for each line and transposing them into velocity space, with respect to their expected lab rest frame energies, with negative values denoting
blueshift.
For the He-like lines, 
velocity profiles were constructed for N\,\textsc{vi}, O\,\textsc{vii} and Ne\,\textsc{ix}, which are the best 
defined He-like complexes (Mg\,\textsc{xi} being poorly constrained in the RGS). 
For the N\,\textsc{vi} and O\,\textsc{vii}  
He-like line complexes, the velocities were centered (at zero velocity) 
on the expected positions of the intercombination emission, as from the 
previous fits it appeared that both of these line profile centroids lie close to these transitions. 
In contrast, the Ne\,\textsc{ix} profile is centered (with zero velocity) 
at the expected position of the forbidden line at 905.1\,eV, as also indicated in the above spectral fits.

Velocity profiles were also created for the H-like (Lyman-$\alpha$) emission lines of N\,\textsc{vii}, O\,\textsc{viii} 
and Mg\,\textsc{xii}, centered at zero velocity on the expected centroid energy of the emission line doublets (the separation 
of which is unresolved in the RGS data). 
The velocity profiles of the three H-like and three He-like profiles are plotted in Figure 7. The profiles have initially been fitted 
with single Gaussians, with the centroid position and $1\sigma$ width (in velocity space) as well as the Gaussian 
normalization (flux) allowed to vary. For the He-like complexes, the 
vertical (dashed-dotted) lines indicate the relative positions in velocity space of the resonance, intercombination and 
forbidden emission line components compared to the line centroids, while for the H-like profiles, the expected center position of the doublet is marked with a vertical line (at zero velocity). The results of these single Gaussian fits to the six line profiles are also tabulated in Table 5. 

The centroid emission of the H-like profiles are generally close to zero velocity and
do not appear to be resolved in terms of their widths compared to the resolution of the RGS spectrometer. 
Indeed, after correcting for the instrumental width due to the response function of the RGS, 
only upper-limits are obtained for the intrinsic velocity widths of the H-like N\,\textsc{vii}, O\,\textsc{viii} and Mg\,\textsc{xii} lines. For instance the limit on the width of the N\,\textsc{vii} profile is $\sigma<290$\,km\,s$^{-1}$
(or FWHM, $<680$\,km\,s$^{-1}$), while for O\,\textsc{viii} the velocity width is restricted to 
$\sigma<470$\,km\,s$^{-1}$ (or FWHM, $<1100$\,km\,s$^{-1}$); see Table 5 for details.
There is an 
marginal indication that the line centroids are slightly blueshifted in the case of N\,\textsc{vii} and O\,\textsc{viii} (with velocities 
of $v_{\rm out} = -410\pm120$\,km\,s$^{-1}$ and $-250\pm130$\,km\,s$^{-1}$ respectively).  
There is also a hint of positive residuals on the blue-wing of the N\,\textsc{vii} line 
(between $-2000$ to $-3000$\,km\,s$^{-1}$) and on the red-wing (near $+2000$\,km\,s$^{-1}$) of the 
O\,\textsc{viii} profile, although neither are statistically significant. Thus although we cannot exclude an underlying broad 
component to either of these profiles, it is likely to be weak. 
Finally for the O\,\textsc{viii} profile, there appears to be 
a marginally significant (at the $\sim 3 \sigma$ level) absorption feature, observed at $E=0.643$\,keV and 
redshifted by $+4500$\,km\,s$^{-1}$ with respect to the O\,\textsc{viii} Ly-$\alpha$ centroid. 
However this may be spurious, as no other absorption lines are associated to any of the 
other profiles at a similar velocity. 

\subsection{The He-like triplets}

In contrast, the He-like profiles shown in Figure 7 all appear significantly broadened when fitted with a single Gaussian 
profile, with intrinsic $1\sigma$ velocity widths of between $\sigma=3200-4000$\,km\,s$^{-1}$ typically (see Table\,5 for the corresponding FWHM values). In comparison the contribution of the width from the instrumental 
resolution of the RGS is negligible at the positions of these lines. The centroid of the Gaussian profile of the N\,\textsc{vi} 
emission is close to that of the intercombination transitions, albeit with a small blueshift of $v_{\rm out} = -1500\pm1000$\,km\,s$^{-1}$. 
Likewise the O\,\textsc{vii} profile is centered at the position of the intercombination emission, with no velocity shift 
required, which is limited to $<730$\,km\,s$^{-1}$. Note that for the broad O\,\textsc{vii} profile to be primarily associated to the forbidden line would instead require a significant blueshift of $-4000$\,km\,s$^{-1}$. However there may still be some 
contribution of the forbidden line to the overall O\,\textsc{vii} profile, as some excess is observed near its expected position in the profile. In contrast the broad 
Ne\,\textsc{ix} Gaussian appears centered at the forbidden line, with any velocity shift of the profile limited to 
$<710$\,km\,s$^{-1}$. 

Next, the He-like profiles were modeled using a combination of the forbidden, intercombination and resonance components, 
with the widths of these components limited to the instrumental RGS resolution, i.e. they are initially assumed to be 
intrinsically narrow and the overall He-like complex consists of a blend of these narrow lines. 
The fit with narrow lines to the profiles are displayed in the left hand panels of Figure 8.
In the case of N\,\textsc{vi}, which is the least well determined of the three He-like line profiles, 
an acceptable fit to the profile of $\chi^{2}/{\rm dof} = 27.0/25$ is obtained from the combination of three narrow components. This is 
comparable to the fit obtained above with a single broad profile, where $\chi^{2}/{\rm dof} = 26.5/28$. 
However in order to model the profile with 3 narrow components, a significant blueshift (of the order $-2000$\,km\,s$^{-1}$) is required, as can be seen in panel (a) in Figure 8 with respect to their expected positions. 
Nonetheless, allowing the intercombination emission 
to have some intrinsic width does improve the fit statistic somewhat to $\chi^{2}/{\rm dof} = 20.7/25$, with an 
intrinsic width of $\sigma_{\rm v}=1840\pm820$\,km\,s$^{-1}$ (see Fig 8, panel b). 

In contrast the O\,\textsc{vii} profile is poorly modeled ($\chi^{2}/{\rm dof} = 92.1/52$) with a combination of 3 narrow lines, as is seen in Figure 8(c), 
where significant flux is left unmodeled by the narrow emission lines. 
Thus allowing a 
broad component (while retaining a possible contribution from any narrow emission if required), 
results in a significant improvement 
in the fit statistic to $\chi^{2}/{\rm dof} = 37.7/51$. The line width then becomes significantly broadened 
as before, with the line centroid consistent with the expected position of the 
intercombination emission. A weak contribution from a narrow forbidden line may still be present 
in the model, as can be seen near $+4000$\,km\,s$^{-1}$ (wrt the intercombination emission) in Figure 8(d), which shows the composite narrow plus broad profile.
Likewise, the Ne\,\textsc{ix} line profile is also poorly modeled with a combination of three narrow line components, 
with $\chi^{2}/{\rm dof} = 73.5/34$ and significant line flux is left on both the red and the blue side of the narrow 
forbidden line (Figure 8, panel e). Thus 
allowing the width of the forbidden line to increase results in a substantial improvement in the fit statistic to 
$\chi^{2}/{\rm dof} = 30.9/33$, while the contribution of any narrow intercombination and resonance lines are negligible 
(Figure 8f). 

It is also possible that the line triplets are composed of a blend of broadened emission from each of the components, 
i.e. with the same velocity width as might be expected if they
originate from within the same zone of gas. 
To test this and to determine the relative contribution of the three components 
towards the overall broad profiles, each He-like profile was fitted with three 
Gaussian components near the expected positions of the forbidden, intercombination and resonance emission lines. 
A common (tied) velocity width was adopted 
between each of the Gaussians. Although the subsequent three Gaussian fits were statistically identical 
to a fit with a single broad Gaussian profile as above, the results give an indication of the relative 
contribution of each component towards the overall profile.
The results of this deconvolution are shown for each triplet in Table\,5. For the O\,\textsc{vii} triplet, the 
emission can be equally composed of contributions from the forbidden, intercombination and resonance emission. 
The line width is then somewhat lower than for a single broad Gaussian, but is still significantly broadened, with 
$\sigma_{\rm v}=2000^{+1200}_{-800}$\,km\,s$^{-1}$. 
In contrast the Ne\,\textsc{ix} broad profile is dominated by the forbidden emission, 
which is not surprising as the overall line profile 
is centered very near the expected position of the forbidden line. The resulting 
common line width in this case is $\sigma_{\rm v}=2300^{+1800}_{-1100}$\,km\,s$^{-1}$ (Table\,5).

Thus for the He-like lines, the broad component appears to dominate these profiles, 
with any narrow line contribution representing a relatively small component of the total line flux. 
The high flux ratio of the intercombination to forbidden emission, for at least the N\,\textsc{vi} and O\,\textsc{vii} triplets, suggests that the electron density of the gas may be relatively high. 
In contrast, the velocity widths of the highest ionization H-like lines are generally narrow or unresolved, while some profiles may have a modest outflow velocity. 

\subsection{Constraints on the Line Emission in the Chandra HETG Spectrum}

In addition to the deep RGS exposure, a 120\,ks exposure was also obtained with Chandra/HETG over the same 
time frame. Given the much shorter HETG exposure and lower area at the lowest energies, 
the signal to noise below 1\,keV does not provide any additional 
constraints on the O\,\textsc{vii} or Ne\,\textsc{ix} line emission. 
Although the primary purpose of the {\it Chandra} exposure was to 
measure the Fe K profile at high resolution, the HETG does provide additional constraints on any possible emission 
from Mg, Si and S above 1\,keV, where in particular its higher resolution compared to the RGS makes the HETG 
more sensitive to any narrow components of the line emission.

The most significant soft X-ray line emission detected in the HETG spectrum arises from the Mg\,\textsc{xi} He-like 
triplet, which is shown in Figure\,9. Indeed at this energy, the resolution of the HEG spectrum is a factor $\times8$ 
higher compared to the RGS. Significant line emission (at $>99.9$\% confidence with $\Delta C=14.8$) 
is revealed in the HEG spectrum just bluewards of the 
expected position of the Mg\,\textsc{xi} forbidden line (at $E_{\rm lab}=1331.1$\,eV), with a measured rest energy of 
$E=1333.5^{+1.3}_{-1.5}$\,eV (see Table 4). 
The equivalent width of the line is $2.2\pm1.0$\,eV, which is consistent with the tentative detection of the 
Mg\,\textsc{xi} line in the RGS spectrum at lower resolution.
Interesting the forbidden line profile appears somewhat broadened in the {\it Chandra} spectrum, compared 
to the HEG FWHM resolution of 320\,km\,s$^{-1}$ at this energy. Indeed the best fit 
line width is $\sigma=2.0^{+1.3}_{-0.8}$\,eV, corresponding to a velocity width of 
$\sigma_{\rm v}=450^{+290}_{-180}$\,km\,s$^{-1}$ (or $1050^{+680}_{-420}$\,km\,s$^{-1}$ at FWHM); see Tables 4 and 5. 
This is consistent with the widths of the narrow H-like components measured above in the RGS, which are largely 
unresolved at lower resolution; 
e.g. O\,\textsc{viii} Lyman-$\alpha$ has a velocity width constrained to $\sigma_{\rm v}<470$\,km\,s$^{-1}$ in the RGS 
spectrum.

Thus the forbidden emission detected and resolved in the {\it Chandra} spectrum appears consistent in origin 
with the narrow line emitting gas revealed in the RGS spectrum. Curiously the Mg\,\textsc{xi} line profile appears slightly 
blue-shifted with respect to the expected position of the forbidden line, with 
an overall blue-shift of $v_{\rm out}=-540^{+340}_{-290}$\,km\,s$^{-1}$. This is also 
consistent with the narrow line profiles observed in the RGS, which may also have a modest blueshift; e.g. 
N\,\textsc{vii} Ly-$\alpha$ with $-410\pm120$\,km\,s$^{-1}$ or O\,\textsc{vii} Ly-$\alpha$ with $-250\pm130$\,km\,s$^{-1}$ 
(see Table 5 and Figure 7). Thus the narrow line gas may be associated to a slight 
outflow velocity, while the small line widths suggest that it is located further from the nucleus than for 
the broad line profiles. 
Note that unlike the detection of the forbidden emission, there is no detection of 
emission from either the intercombination 
or resonance components of the Mg\,\textsc{xi} line triplet (see Figure 9), with upper limits of $<0.4$\,eV and $<1.5$\,eV 
on the equivalent widths respectively. This suggests that the narrow lined emission originates primarily from 
lower density gas (given the dominance of the forbidden over the intercombination emission); 
the line diagnostics from the triplets will be discussed further in Section\,6.

Aside from at Mg\,\textsc{xi} (and also at the iron K$\alpha$ line, which will be presented by Nardini et al. 2016), 
there are no other significant detections of line emission in the HETG spectrum. There are some weak 
indications of emission associated with the forbidden lines of Si\,\textsc{xiii} and S\,\textsc{xv}, 
although neither of these are formally significant (with $\Delta C=2$ and $\Delta C=5$), 
with upper limits of $<1.1$\,eV and $<2.5$\,eV respectively on their equivalent widths. 
Neither are any of the H-like lines from Mg, Si or S detected. 
Generally this is expected given the continuum dominated nature of Ark\,120, while it is the exceptionally 
deep RGS exposure that has enabled the detection of the soft X-ray line emission for the first time from this AGN. 

\subsection{Is there any Warm Absorption towards Ark\,120?}

In addition to the emission, here we place limits on any intrinsic ionized absorption towards Ark\,120. 
We included a Solar 
abundance \textsc{xstar} 
multiplicative table of photoionized absorption spectra in the spectral model fitted to the RGS
at the rest frame of the AGN, while 
the outflow velocity was allowed to vary between $\pm3000$\,km\,s$^{-1}$ to allow for any velocity shift. 
A turbulence velocity of 
$\sigma=300$\,km\,s$^{-1}$ was chosen, to account for any narrow absorption lines.
The ionization parameter of the absorption table was varied between 
$\log\xi=0-3$\footnote{The ionization parameter is defined here as 
$\xi = L_{\rm ion}/nr^{2}$, where $L_{\rm ion}$ is the ionizing luminosity measured between 1--1000 Rydberg, 
$n$ is the electron density and $r$ is the radial distance from the X-ray source. The units of the ionization 
parameter are erg\,cm\,s$^{-1}$.} 
to allow for the typical range in ionization seen towards other Seyfert 1s, with prominent warm 
absorption components 
(Kaspi et al. 2002, Blustin et al. 2005, Detmers et al. 2011, Reeves et al. 2013, Di Gesu et al. 2015). 
The total $N_{\rm H}$ column was recorded for each value of the ionization parameter, 
in increments of 0.1 in $\log \xi$ space. Tight limits were placed on the total absorbing column density, 
of between $N_{\rm H}<1.8\times10^{19}$\,cm$^{-2}$ and $N_{\rm H}<3.4\times10^{19}$\,cm$^{-2}$ over the ionization 
range of $\logxi=0-2$, with a less stringent constraint towards higher ionization values
($N_{\rm H}<6.0\times10^{20}$\,cm$^{-2}$ for $\logxi=3$). Furthermore the limits on the total O column, 
over the ionization range $\logxi=0-2$ (where typically most of the absorption due to O\,\textsc{vii} and 
O\,\textsc{viii} is expected to occur), varies between $N_{\rm O}<1.2\times10^{16}$\,cm$^{-2}$ and 
$N_{\rm O}<3.4\times10^{16}$\,cm$^{-2}$. 

Indeed upon inspection of the RGS spectra (Figure 6) there are no apparent absorption features 
at positions close to the expected resonance transitions of the strongest lines in the soft X-ray band. 
The findings and constraints on the column density of the absorption 
here are consistent with those of Vaughan et al. (2004), who published their findings 
from a single orbit (112\,ks RGS exposure) XMM-Newton observation taken in 2003 and whom also found no evidence 
for any intrinsic X-ray absorption towards Ark\,120. Furthermore there is no known UV absorber associated 
with Ark\,120 (e.g. Crenshaw et al. 1999, Crenshaw \& Kraemer 2001). 
This confirms Ark\,120 as the prototype bare Seyfert 1 AGN, with no known intrinsic absorption along the 
line of sight towards the AGN.

\section{Discussion}\label{sec:discussion}

\subsection{The Origins of the Ionized Soft X-ray Emission in Ark\,120}

While there is no intrinsic X-ray absorption towards Ark 120 (aside from the 
neutral ISM absorption due to the Milkyway), the high signal to noise RGS 
spectrum has revealed several broad and narrow emission line profiles 
associated to the AGN. As shown in Section\,5, these broad profiles are associated with the He-like emission from N\,\textsc{vi}, O\,\textsc{vii} and 
Ne\,\textsc{ix} with velocity widths in the range from 
4000--8000\,km\,s$^{-1}$ (FWHM). In addition, the line profile 
modeling in Section\,5.3 appears to exclude 
the possibility of the He-like triplets 
being composed of purely a blend of narrow unresolved emission lines from the forbidden, intercombination and resonance 
components, with the intrinsic velocity widths being resolved by the RGS (see Figure 8).
In contrast the H-like profiles from 
N\,\textsc{vii}, O\,\textsc{viii} and Mg\,\textsc{xii} are narrow, 
for instance for the O\,\textsc{viii} Ly$\alpha$ emission the upper limit on the 
Gaussian velocity width is $\sigma<470$\,km\,s$^{-1}$.

Thus while there is no direct absorption along the line of sight 
associated to the AGN (see also 
Vaughan et al. 2004, Matt et al. 2014), 
the fact that there is significant soft X-ray line emission associated with  
Ark\,120 suggests the AGN is not intrinsically bare. Thus the lack of absorption 
may just indicate that we are viewing the AGN along a preferential 
line of sight, with relatively little ionized gas along our direct view. 
The detection of narrow soft X-ray emission from many Seyfert 
galaxies, associated with photoionized or photoexcited gas, 
has proven common from grating observations of obscured AGN (Kinkhabwala et al. 2002), 
with the origin likely to be on scales consistent with the 
AGN Narrow Line Region (NLR) and perhaps arising from a large scale outflow. 
However recent observations are now also revealing the detections 
of broad soft X-ray line profiles from several Seyfert 1 galaxies, 
with velocity widths of several thousand km/s, suggesting an 
origin commensurate with the AGN Broad Line Region (BLR). Examples include NGC\,4051 
(Ogle et al. 2004, Pounds \& Vaughan 2011), Mrk\,335 (Longinotti et al. 2008), 
Mrk\,841 (Longinotti et al. 2010), 3C\,445 (Reeves et al. 2010),  Mrk\,509 (Detmers et al. 2011), MR\,2251-178 (Reeves et al. 2013) and NGC\,5548 (Kaastra et al. 2014). 

Indeed another AGN that bares some similarity to Ark\,120 is the bare Seyfert 1 galaxy, Mrk\,590. 
Here \xmm\ and \chandra\ showed no intrinsic absorption, but the presence of narrow ionized emission associated 
with highly ionized iron (Fe\,\textsc{xxv} and Fe\,\textsc{xxvi}) as well as from ionized Oxygen (O\,\textsc{viii}); 
see Longinotti et al. (2007). In addition, the extended soft X-ray emission detected in the {\it Chandra} image of Mrk\,590  
implies the presence of ionized gas on larger (kpc) scales.
The overall picture may be similar to what is observed in 
Ark\,120, namely that little ionized gas is detected along the line of sight, but evidence for photoionized gas 
is still seen from the circumnuclear gas out of the direct view. In 
Ark\,120, the detection of both narrow and broad lines implies the 
existence of soft X-ray emitting gas over a wide range of spatial scales, 
from the sub-pc BLR gas out to the more distant NLR, with the narrow highly ionized H-like emission possibly 
associated to the warm scattering gas on larger scales. 
In the following section, we attempt to quantify the location and physical properties of 
the soft X-ray emitting gas in Ark\,120.

\subsection{Constraints from the He-like line triplets}

The lack of intrinsic X-ray absorption towards Ark\,120 allowed a clean measurement of the emission from 
the He-like triplets.
Given the constraints on the He-like line triplets, we can place an estimate on the 
density and subsequently infer the likely radial location of the emitting gas. The line ratios $G = (x + y + z) / w$ and 
$R = z / (x + y)$ give a measure of the temperature and density of the gas, where $z$ corresponds 
to the forbidden line, $(x + y)$ to the intercombination emission and $w$ to the resonance line 
(Porquet \& Dubau 2000). Taking the ratios of the fluxes of the line components 
measured from the O\,\textsc{vii} triplet (Table\,5), then $R=1.1\pm0.6$, resulting from a relatively 
equal contribution of the broad forbidden and intercombination components.
This indicates that the gas is of relatively high density. From the calculations of 
Porquet \& Dubau (2000), this ratio corresponds to an electron density of $n_{\rm e}\sim 10^{11}$\,cm$^{-3}$. 
On the other hand the high G ratio, of $G=2.1\pm1.0$, 
indicates that the gas being photoionized rather than collisionally ionized, with 
a temperature of $T<10^6$\,K. Note that photoionization rather than photo-excitation appears
the more dominant mechanism in Ark\,120, given the relative weakness of the higher order emission lines 
(see Fig 5, Kinkhabwala et al. 2002); for instance for O\,\textsc{vii} the ratio of the He$\alpha$ to 
He$\beta$ emission is $>10$.

In contrast for the He-like triplet of Ne\,\textsc{ix}, the forbidden line appears to dominate over any intercombination 
emission (see Table 5), as generally the Ne\,\textsc{ix} triplet is more sensitive towards higher densities 
compared to O\,\textsc{vii}. 
Thus the R ratio of $R>3.3$ implies a limiting density of $n_{\rm e}\ls 2\times10^{11}$\,cm$^{-3}$. 
So for Ark\,120, the broad lined soft X-ray emitting gas appears consistent with a density in the 
range from $10^{11}\ls n_{\rm e}\gs 2\times10^{11}$\,cm$^{-3}$. The G ratio from the Ne\,\textsc{ix} triplet is 
$G>3.8$, consistent with a photoionized plasma.
Note that the constraints on $R$ obtained from the N\,\textsc{vi} triplet are consistent with this, 
with a resulting lower limit on the density of $n_{\rm e}>10^{10}$\,cm$^{-3}$.

\subsection{The Ionization State and Location of the Gas}
 
Thus the above density would seem to imply an origin of the broad 
line emission consistent with the optical BLR (Davidson \& Netzer 1979). 
The ionization of the emitter can also be constrained, 
given the measured line flux ratio of O\,\textsc{vii} He$\alpha$\,/\,O\,\textsc{viii} Ly$\alpha$\,$\sim10$, 
e.g. Table~4. Indeed from fitting an \textsc{xstar} emission model with a density of 
$n_{\rm e}=10^{11}$\,cm$^{-3}$ to model the O\,\textsc{vii} emission (see Section\,6.4), 
an ionization parameter of $\logxi=0.5$ is required in order for the less ionized He-like emission 
to dominate over the H-like emission.

From this an estimate of the radial distance of the emitter can be obtained via the definition of the ionization parameter, 
i.e. $r = (L_{\rm ion} / \xi n_{\rm e})^{1/2}$, where $L_{\rm ion}$ is the $1-1000$\,Rydberg luminosity  
and $n_{\rm e}$ is the electron density. From the broad-band UV to hard X-ray SED, the 
ionizing luminosity of Ark\,120 is 
$L_{\rm ion}\sim 10^{45}$\,erg\,s$^{-1}$ (Porquet et al. 2016).
If we adopt a density of $n_{\rm e}=10^{11}$\,cm$^{-3}$ from the above considerations, 
then the emitting radius is $r=5\times10^{16}$\,cm 
(or $\sim0.01$\,pc), again 
consistent with typical BLR radii (Kaspi et al. 2005). Furthermore the distance to the optical BLR, 
inferred from time delays of the H$\beta$ line wrt to the continuum in Ark\,120 is $\tau\sim40$\,light days 
(Peterson \& Gaskell 1991, Wandel, Peterson \& Malkan 1999), 
equivalent to a distance of $10^{17}$\,cm.  
The radius of the emission can also be estimated from the O\,\textsc{vii} or Ne\,\textsc{ix} velocity 
widths of $\sigma\sim2000$\,km\,s$^{-1}$ or a FWHM of $\sim5000$\,km\,s$^{-1}$, adopting the 
more conservative lower value from the triplet deconstruction (see Table 5). Assuming 
a standard virial relation between the black hole mass and the radius $r$, of $3\sigma^{2}=GM/r$ 
and adopting a black hole mass of $1.5\pm0.2\times10^{8}$\Msun\ 
for Ark\,120 from reverberation mapping (Peterson et al. 2004), 
gives a radius of $r\sim10^{17}$\,cm, consistent with the 
above estimates.

In comparison the optical H$\beta$ FWHM line width of Ark\,120 is $5850\pm480$\,km\,s$^{-1}$ 
(Wandel, Peterson \& Malkan 1999), 
which is similar to (or slightly smaller) than the typical X-ray broad line widths measured here. 
Furthermore the core of the 6.4\,keV iron K$\alpha$ emission line is also resolved in the simultaneous Chandra/HETG 
spectrum, with a FWHM width of $4500^{+2500}_{-1500}$\,km\,s$^{-1}$ (Nardini et al. 2016), 
consistent with the above estimates. 
Thus the observations suggest that the soft X-ray 
broad emission lines originating from a higher ionization phase of the AGN BLR, with radii in 
the typical range from $5\times10^{16}-10^{17}$\,cm. 

In contrast the H-like profiles measured in the spectra from N, O and Mg appear narrow and are unresolved 
by the RGS -- see Figure 7 and Table 5. The constraints on the density of the narrow lined emitting gas are not 
as tight as for the broad lines, with the Mg\,\textsc{xi} triplet measured from {\it Chandra} 
(forbidden dominating over intercombination with $R>3.2$) indicating $n_{\rm e}<10^{12}$\,cm$^{-3}$. 
Thus although we cannot directly measure the density of the narrow line emitting gas 
from line diagnostics, we can calculate its likely radial location from the limits on the velocity widths. 
If we take the line width for the resolved Mg\,\textsc{xi} forbidden line from the {\it Chandra} spectrum, 
with $\sigma=450^{+290}_{-180}$\,km\,s$^{-1}$, then the likely radius of the emitting gas is on 
pc scales. Note this is also consistent with the limits on the widths of the narrow lines seen in the RGS, 
e.g. for the O\,VIII Ly$\alpha$ line then $\sigma<470$\,km\,s$^{-1}$. 
For comparison, the expected distances of the
torus and of the Narrow Line Region (NLR) are about 3\,pc and
100\,pc, using the following formula of Krolik \& Kriss (2001) and Mor, Netzer \& Elitzur (2009)
respectively:

\begin{equation} 
R_{\rm torus}\sim L_{\rm ion,44}^{1/2} ~~~~~~~~ (pc)
\end{equation}

\begin{equation} 
R_{\rm NLR} = 295 \times L_{46}^{0.47\pm 0.13} ~~~~ (pc).
\end{equation}

Thus, unlike for the broad line emitting gas, the narrow line X-ray emission appears consistent with radial locations 
commensurate with the pc scale torus or innermost NLR. This is also at a similar location to some of the soft X-ray 
warm absorbers inferred in Seyfert 1 AGN (see Tombesi et al. 2013 and references therein), 
which may imply that we are viewing similar ionized gas in Ark\,120 out of our direct line of sight.

\subsection{The Covering Fraction and Geometry of the Gas}

The luminosity of the soft X-ray line emission can also be used to calculate 
the global covering factor of the gas (see also Nucita et al. 2010). From the photoionization modeling, 
the normalization (or flux), $\kappa_{\rm xstar}$, of an emission component is defined 
by \textsc{xstar} (Kallman et al. 2004) in terms of:
\begin{equation}
\kappa_{\rm xstar} = f\frac{L_{38}}{D_{\rm kpc}^2}
\end{equation}
where $L_{38}$ is the ionizing luminosity in units of $10^{38}$\,erg\,s$^{-1}$, 
$D_{\rm kpc}$ is the distance to the AGN in kpc. Here $f$ is the covering fraction 
of the gas with respect to the total solid angle, where $f = \Omega / 4\pi$. For a 
spherical shell of gas, $f=1$, while $L$ is the quasar ionizing luminosity that illuminates 
the photoionized shell. 
The ionizing luminosity obtained from the best fit spectral 
model and corrected for the intervening Galactic absorption, is $L=4.5\times10^{44}$\,erg\,s$^{-1}$ 
over the 1--1000\,Rydberg band\footnote{Note that the ionizing luminosity returned from a broad-band fit to Ark\,120 
with the \xmm\ EPIC and OM data, obtained from the \textsc{optxagn} disk + corona 
model (Porquet et al. 2016, in preparation), is $\sim 10^{45}$\,erg\,s$^{-1}$.}.
Thus for Ark\,120 with $D=133$\,Mpc and for a spherical shell, 
the expected {\sc xstar} normalization from above is $\kappa_{\rm xstar}=2.5\times10^{-4}$. 
Hence for a given column density of gas, this sets the total luminosity of the soft X-ray photoionized emission.

As a first step, model emission line spectra were then generated with this overall normalization factor 
within \textsc{xstar}, 
which gives the predicted emission originating from a fully covering spherical shell of gas (with $f=1$), 
illuminated by an AGN ionizing luminosity of $L$. The ionization of the gas was fixed at $\logxi=0.5$, consistent 
with the \textsc{xstar} fits to the lines (see below). 
Taking as an example the case of the strong O\,\textsc{vii} He$\alpha$ broad emission line observed 
in the RGS spectrum, we then compared the 
observed line luminosity to that predicted by the \textsc{xstar} model and then used the ratio of the 
observed to predicted line luminosity to calculate the global covering fraction of the gas. 
From the Ark\,120 spectrum, the observed luminosity of the broad O\,\textsc{vii} emission is 
$L_{\rm OVII}=4.0\pm1.2\times10^{41}$\,erg\,s$^{-1}$. 
In comparison, for a column density of $N_{\rm H}=10^{21}$\,cm$^{-2}$, the O\,\textsc{vii} luminosity 
predicted by the \textsc{xstar} model for a fully covering shell of gas 
is $1.2\times10^{42}$\,erg\,s$^{-1}$. The ratio of the observed to predicted luminosity 
then gives the geometric covering fraction of the emitter, which for $N_{\rm H}=10^{21}$\,cm$^{-2}$ is $f=0.33\pm0.10$. 
Similarly, for a higher column of $N_{\rm H}=10^{22}$\,cm$^{-2}$, the predicted O\,\textsc{vii} 
luminosity of a spherical shell is higher, with $3.6\times10^{42}$\,erg\,s$^{-1}$ and thus the covering fraction 
is then lower, with $f=0.11\pm0.03$.  
 
To provide a more quantitative estimate of the covering fraction, 
the RGS spectrum was then fitted with the \textsc{xstar} emission 
models. In order to reproduce the strongest emission lines present in the Ark\,120 spectrum, three different 
photoionized emission zones were required and their properties are summarized in Table\,6.
Two of these emission zones appear to be broadened 
(with $\sigma=3000$\,km\,s$^{-1}$); these are required to model (i) the broad O\,\textsc{vii} emission 
(as well as at N\,\textsc{vi}) and (ii) the broad Ne\,\textsc{ix} emission, with the lower ionization zone 
(with $\log \xi=0.5\pm0.1$) responsible for O\,\textsc{vii}. A third, highest ionization zone ($\log \xi=2.3\pm0.4$) 
with a narrow velocity width ($\sigma=300$\,km\,s$^{-1}$) is responsible for the narrow H-like lines, such 
as from O\,\textsc{viii} Ly-$\alpha$.
The column density of the gas in emission is not known a priori (due to the lack of absorbing gas) and is highly degenerate with the emission normalization. Thus instead of directly fitting both the column and the normalization 
of the \textsc{xstar} emission zones, the column density was varied over the range $3\times10^{20}<N_{\rm H}<10^{22}$\,cm$^{-2}$, adopting 8 different values and at each fixed $N_{\rm H}$ value the spectrum was refitted to obtain the normalization of the \textsc{xstar} component. Thus by comparing the observed normalization ($\kappa_{\rm obs}$)
of a fitted emission component with the predicted value from equation\,3 ($\kappa_{\rm xstar}$), 
the covering fraction for a given column can be calculated by $f=\kappa_{\rm obs}/\kappa_{\rm xstar}$.

The resulting plot of covering fraction vs. column for the O\,\textsc{vii} emission zone is shown in Figure\,10, 
noting that similar results are also found for the Ne\,\textsc{ix} zone and the (narrow) O\,\textsc{viii} zone 
(see Table 6).
The overall trend is for the covering fraction of the gas to decrease with increasing column. 
This would be expected as increasing the column density of the gas clouds increases their soft X-ray luminosity, 
requiring the overall covering fraction to decrease in order to compensate. A 
minimum column density of $N_{\rm H}>3\times10^{20}$\,cm$^{-2}$ is required to reproduce the required 
O\,\textsc{vii} luminosity, if the gas is fully covering the AGN with $f=1$. This fully covering 
scenario appears less likely, as the upper-limit to the line of sight column of soft X-ray warm absorbing gas 
towards Ark\,120 is at least factor of $\times10$ lower; with $N_{\rm H}<3\times10^{19}$\,cm$^{-2}$ for $\logxi=1$, 
see Section 5.5. Instead the covering fraction is likely to be lower, with no significant distribution of gas along the 
line of sight, allowing the column density out 
of the direct line of sight to be higher. Indeed once the column density approaches
$N_{\rm H}=10^{22}$\,cm$^{-2}$, then the covering reaches a limiting value of $f=0.1$. 
This corresponds to a likely {\it minimum} covering fraction of gas 
as increasing the column density above $N_{\rm H}>10^{22}$\,cm$^{-2}$ 
has little effect on the total line luminosity; 
i.e at these columns and higher, the emitting clouds become optically 
thick at soft X-rays, with little change in the resulting soft X-ray line luminosity.

\subsection{X-ray Broad and Narrow Line Regions}

Thus the above calculations, in order to reproduce the broad soft X-ray line emission, requires 
the emitting gas to have a typical column of up to $N_{\rm H}\sim10^{22}$\,cm$^{-2}$, with a covering fraction of 
at least 10\% of $4\pi$ steradian and for the gas to lie out of the direct line of sight. 
The mass of the emitting gas implied from these calculations is $\sim 0.1$\,M$_{\odot}$, 
for $N_{\rm H}=10^{22}$\,cm$^{-2}$ and $f=0.1$. 
Overall this scenario is 
similar to what is usually inferred from studies of the optical--UV BLR (Baldwin et al. 1995, Gaskell 2009), 
with the distribution of emitting BLR clouds likely non-spherical and following a flattened or disk-like geometry 
(Krolik et al. 1991, Eracleous \& Halpern 2003). 
The latter distribution of emitting clouds can also account for the lack of absorption towards 
Ark\,120 if the gas lies out of the direct line of sight and our view of the AGN is relatively pole-on compared 
to the plane of the disk. This may also be consistent with the morphology of the 
host spiral galaxy being relatively pole-on in Ark\,120 
(Nordgren et al. 1995). The opposite may be the case in AGN where variable X-ray obscuration events occur, 
when the line of sight can be intercepted by compact clouds consistent with observed BLR distances. 
Such X-ray obscuration events have been observed several Seyfert galaxies, such as in Mrk 335 (Longinotti et al. 2013), NGC 3227 (Lamer et al. 2003), NGC 1365 (Risaliti et al. 2005b, Braito et al. 2014), NGC 4051 (Terashima et al. 2009, Lobban et al. 2011), NGC\,3516 (Turner et al. 2008) and NGC 5548 (Kaastra et al. 2014). 

In contrast to the broadened emission, the kinematics of the narrow soft X-ray emission lines are consistent 
with arising from pc scale distances or higher. 
The typical column density and covering fraction of the narrow emission line gas is similar to 
broad lined gas; for a column of $N_{\rm H}=1\times10^{21}$\,cm$^{-2}$, the covering fraction is 
consistent with $f=1$, while the covering fraction decreases to $\sim 10$\% for $N_{\rm H}=1\times10^{22}$\,cm$^{-2}$.
Thus the derived columns and covering fractions are thus likely 
too low to be commensurate with emission from a Compton thick pc-scale torus. However the origin of the gas in emission  
may arise from the AGN Narrow Lined Region (NLR) and 
is in general agreement with the typical columns and spatial locations of matter 
inferred along the line of sight in the warm absorbers that are observed in many other Seyfert 1 AGN (e.g. Mrk 509, 
Kaastra et al. 2012). Indeed the kinematics of the narrow emission lines, in terms of the velocity widths and 
a tentative indication of modest blueshift (both of the order of a few hundred km\,s$^{-1}$), 
may imply we are observing emission from a pc scale outflow, but viewed 
out of the line of sight towards Ark\,120.


\section{Acknowledgements}

J.\,N.\,Reeves acknowledges financial support via {\it Chandra} grant number GO4-15092X. 
J.\,N.\,Reeves and T.\,J.\,Turner both acknowledge support from 
NASA grant NNX15AF12G. 
D. Porquet acknowledges financial support from the French GDR PCHE and from  the European Union Seventh Framework Program
(FP7/2007-2013) under grant agreement number 312789.
J.\,N.\,Reeves, E.\,Nardini, A.\,Lobban also acknowledge the financial support of the STFC.
This research is based on observations obtained with the XMM-Newton, and ESA science
mission with instruments and contributions directly funded by ESA
member states and the USA (NASA) and on observations made by the {\it Chandra} X-ray Observatory.
This research has made use of the CIAO software provided by the Chandra X-ray Center (CXC).


\clearpage

\begin{figure}
\begin{center}
\rotatebox{-90}{\includegraphics[width=11cm]{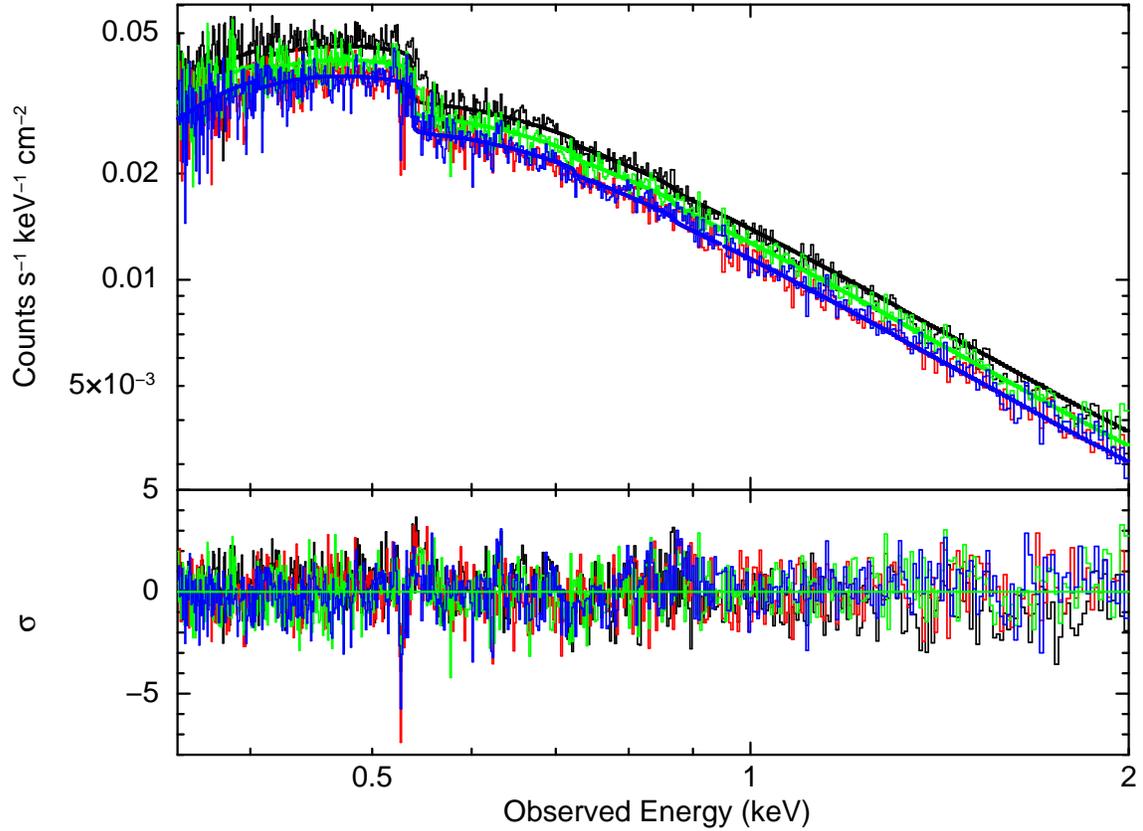}}
\end{center}
\caption{The upper panel shows the XMM-Newton RGS count rate spectra, divided by the 
instrumental effective area. The 4 individual RGS sequences in 2014 are plotted 
as histograms (in black, red, green and blue respectively). The baseline continuum 
model, as described in the text, is shown as a solid line. The lower 
panel show the residuals due to the model, the deviation seen near to 0.52\,keV 
is due to bound-bound absorption from neutral O\,\textsc{i} due to our own Galaxy. 
Note all 4 RGS spectra are consistent with one another, except for a $\pm10$\% 
offset due to the different flux levels of each observation.} 
\label{rgs4spectra}
\end{figure}
\clearpage

\begin{figure}
\begin{center}
\rotatebox{-90}{\includegraphics[width=4.5cm]{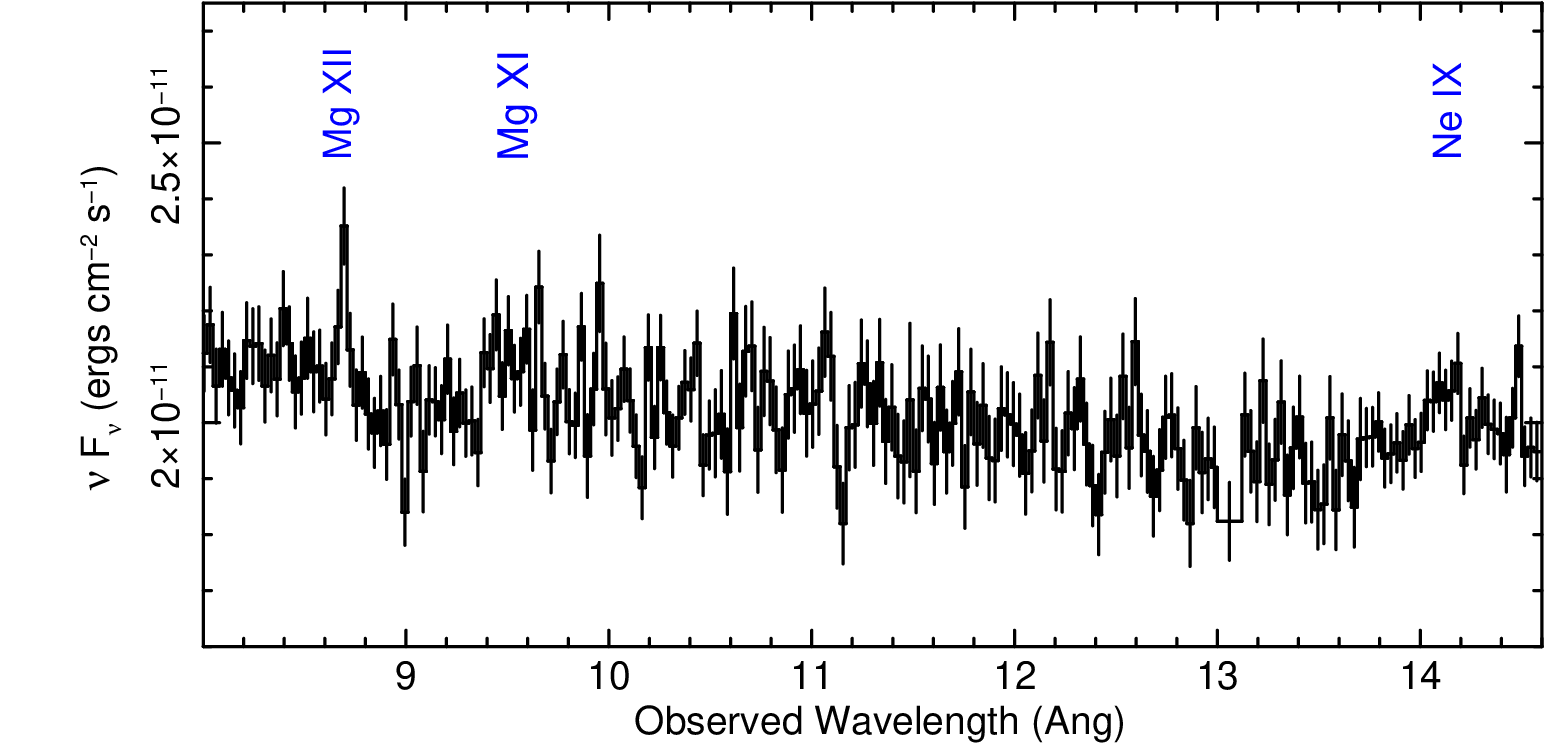}}
\rotatebox{-90}{\includegraphics[width=4.5cm]{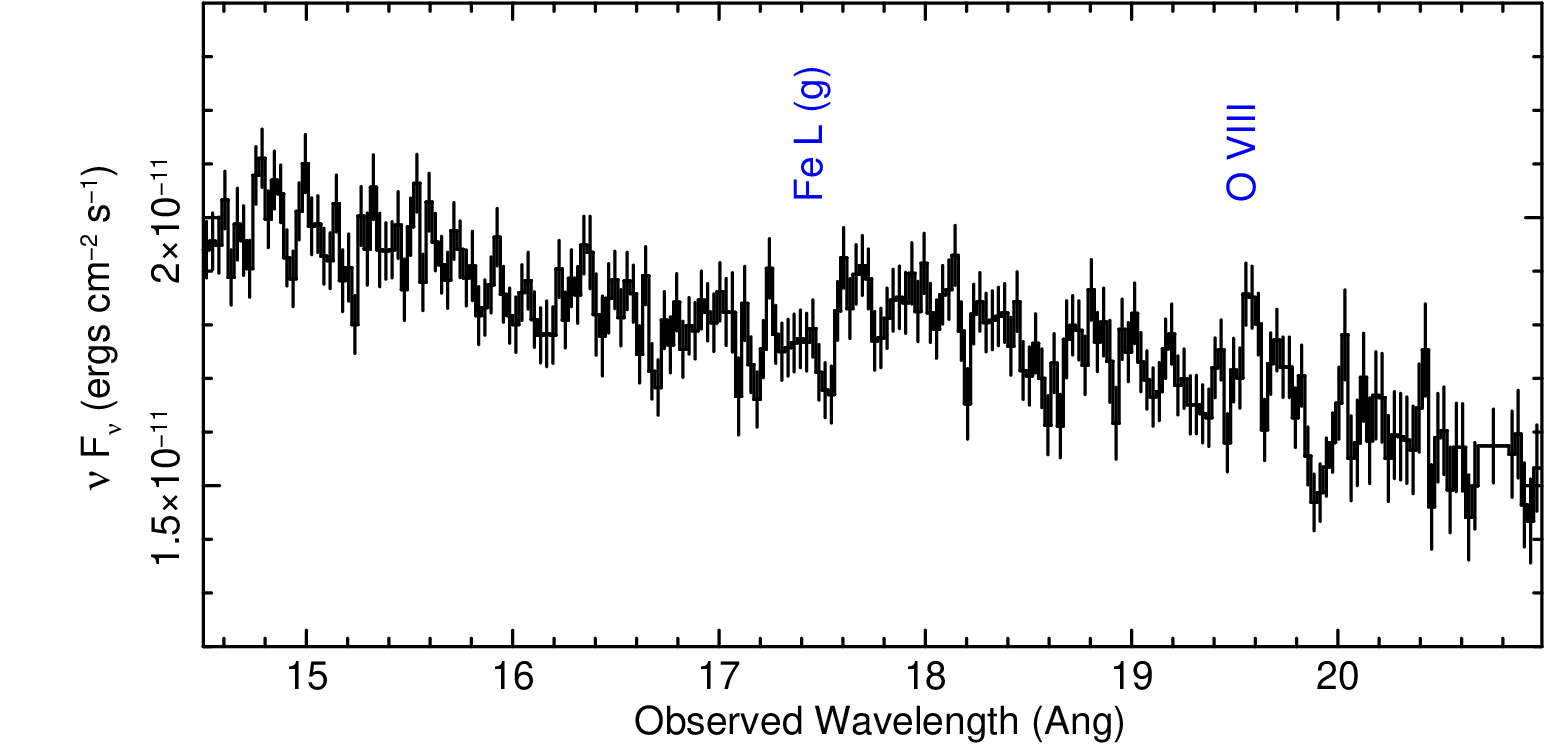}}
\rotatebox{-90}{\includegraphics[width=4.5cm]{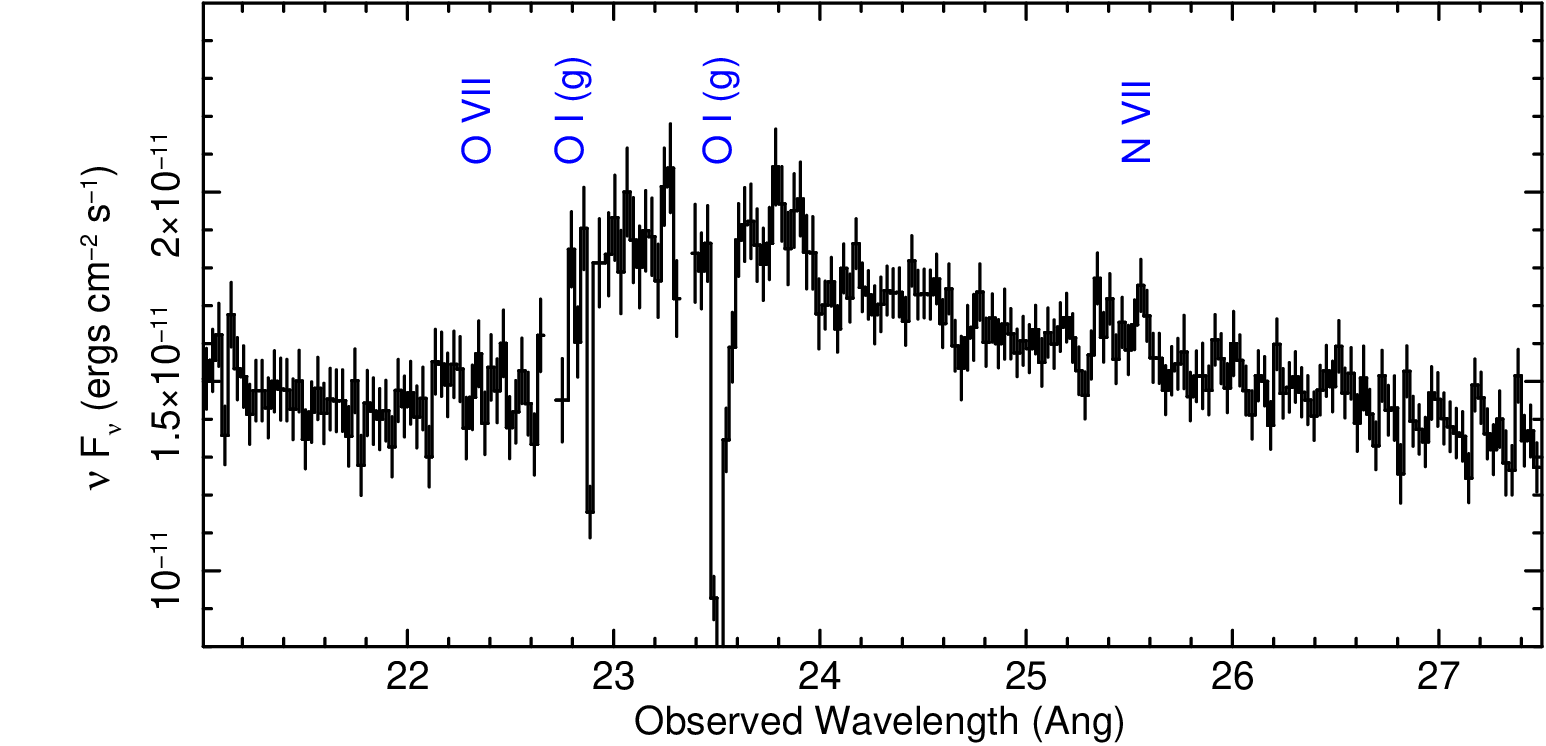}}
\rotatebox{-90}{\includegraphics[width=4.5cm]{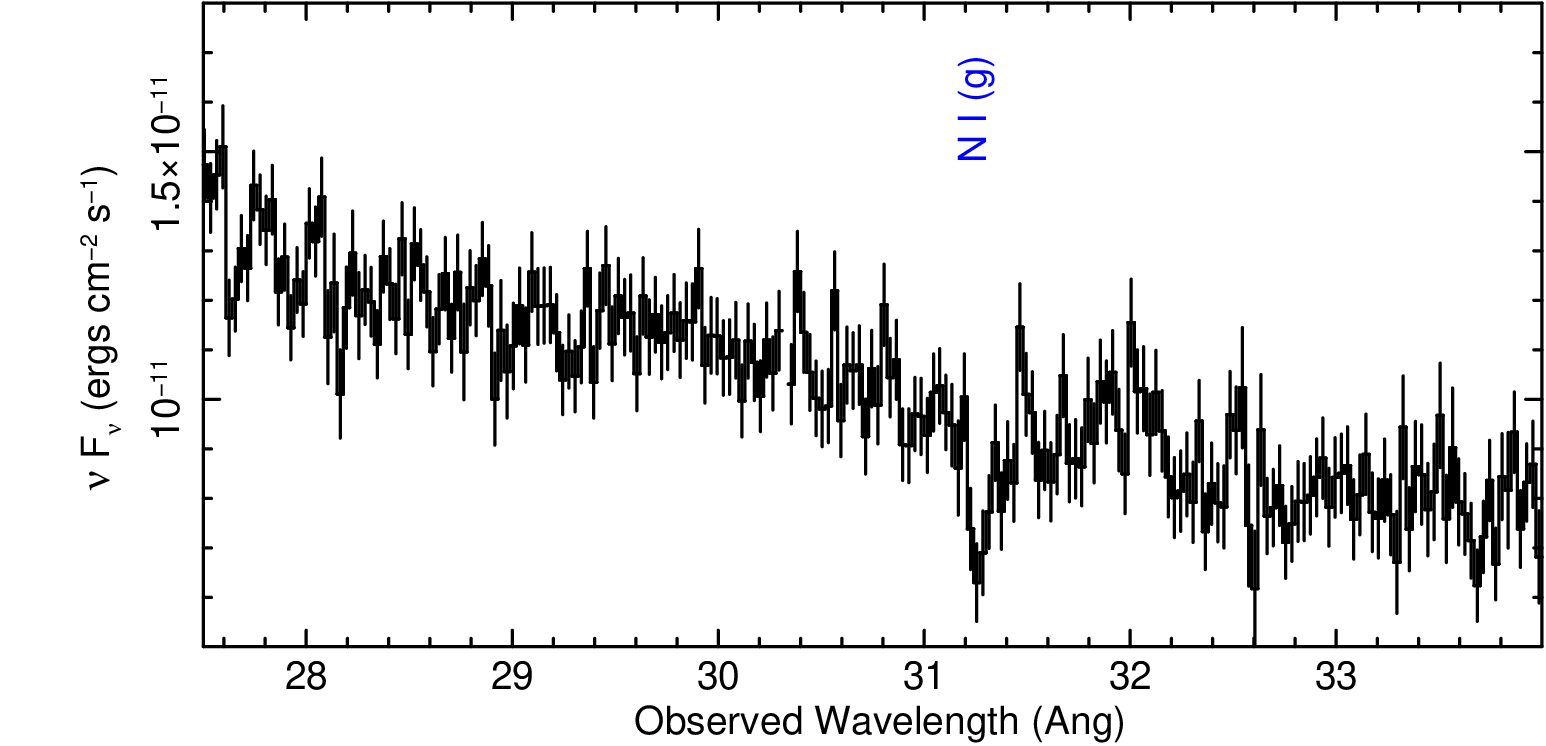}}
\end{center}
\caption{The four panels show the fluxed 2014 RGS spectrum of Ark\,120, obtained from 
combining all 4 sequences and from the RGS\,1 and RGS\,2 data. The spectra have been 
fluxed against a power-law continuum of photon index $\Gamma=2$ to create a $\nu F_{\nu}$ 
plot and are plotted against wavelength (in Angstroms) in the observed frame. The wavelengths 
of some of the expected emission lines from abundant elements are indicated, while the 
predicted position of absorption due to the ISM of our own Galaxy are also marked 
(denoted with a $g$ symbol). Note the strong absorption lines and edge due to neutral O and N 
from our own Galaxy, as well as the possible presence due to photoionized emission lines 
from N, O, Ne and Mg in the rest-frame of Ark\,120. }
\label{rgs-panels1}
\end{figure}

\begin{figure}
\begin{center}
\rotatebox{-90}{\includegraphics[width=8cm]{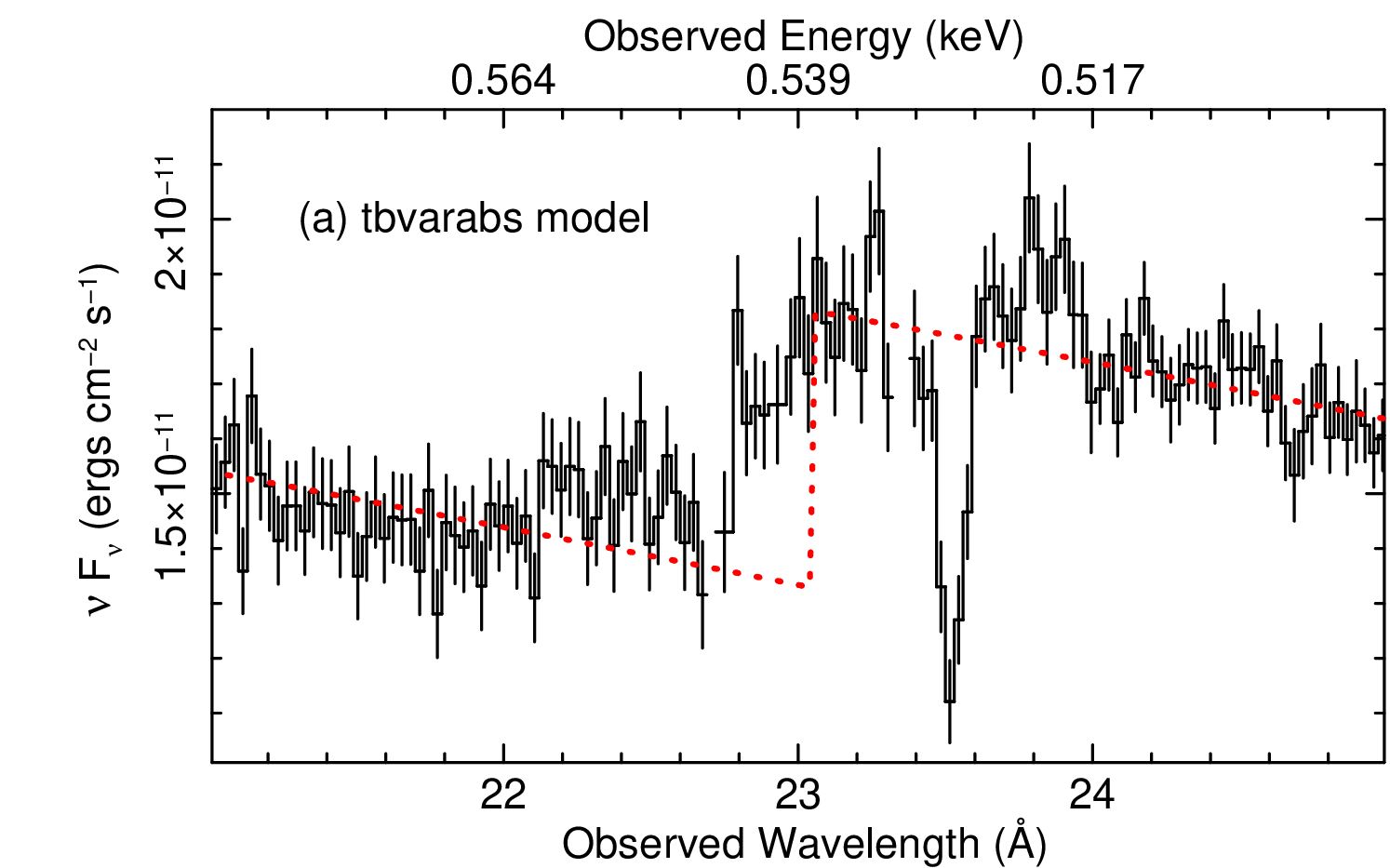}}
\rotatebox{-90}{\includegraphics[width=8cm]{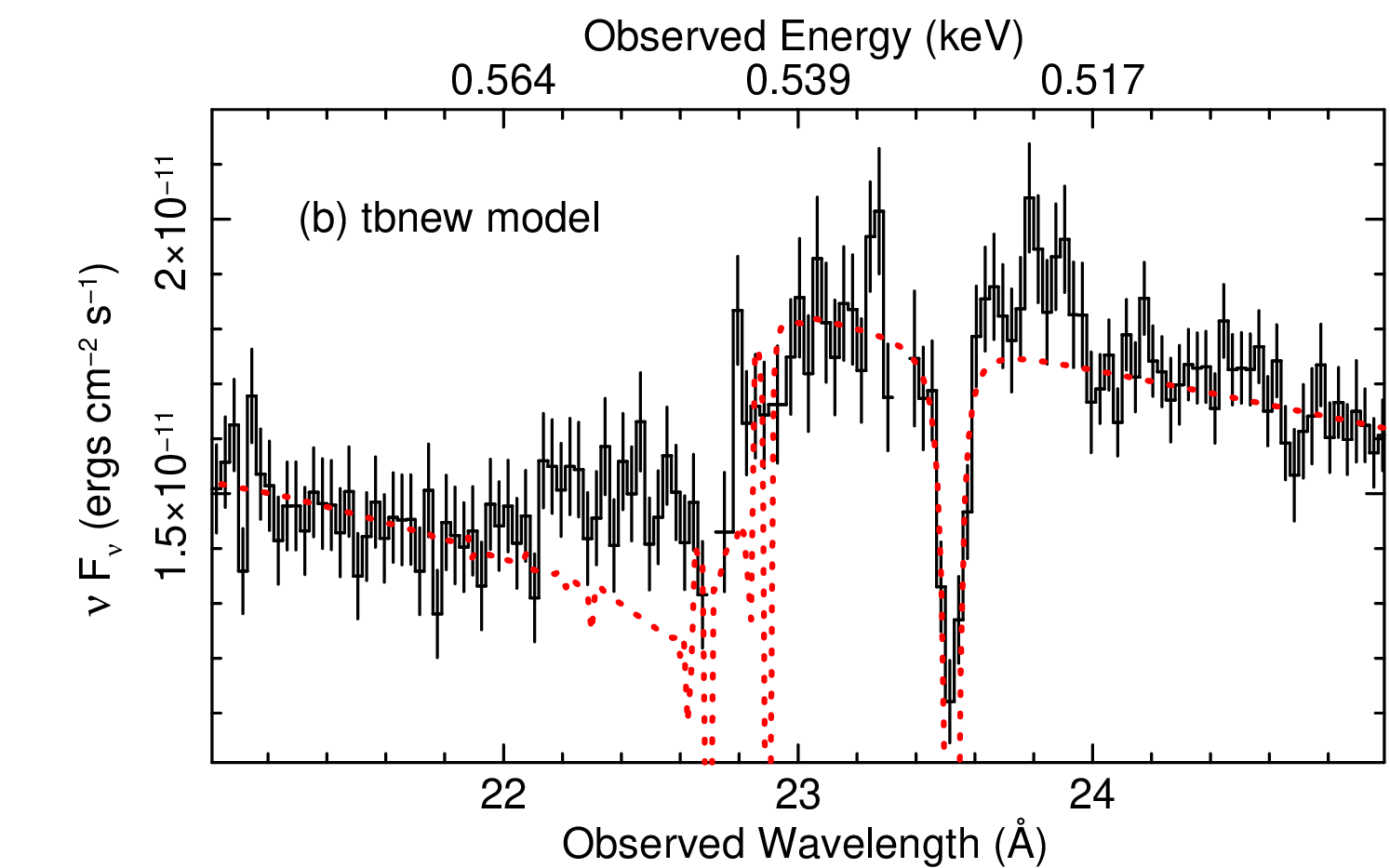}}
\end{center}
\caption{A zoom in of the fluxed RGS spectrum of Ark\,120 in the neutral O K edge region, due to absorption in the Galactic ISM towards Ark\,120. 
Wavelength (energy) is plotted in the observed ($z=0$) frame.
The top panel (a) shows the data compared against the \textsc{tbvarabs} absorption model (red dotted line), where a simple O\,\textsc{i} edge 
which is not able to correctly account for the absorption in this region. The lower panel (b) shows the comparison against the \textsc{tbnew} 
model, which reproduces well the O\,\textsc{i} K$\alpha$ absorption line at 23.5\AA\ (or 527\,eV), while the 
drop around the edge below 23\AA\ is more gradual due to the higher order resonances present in the model. The excess blue-wards of the 
O edge is likely due to O\,\textsc{vii} emission associated with the AGN. Note that the models are superimposed on the spectra 
that have been fluxed against a $\Gamma=2$ powerlaw, but the data are not unfolded against either absorption model.}
\label{tbnew}
\end{figure}

\clearpage

\begin{figure}
\begin{center}
\rotatebox{-90}{\includegraphics[width=11cm]{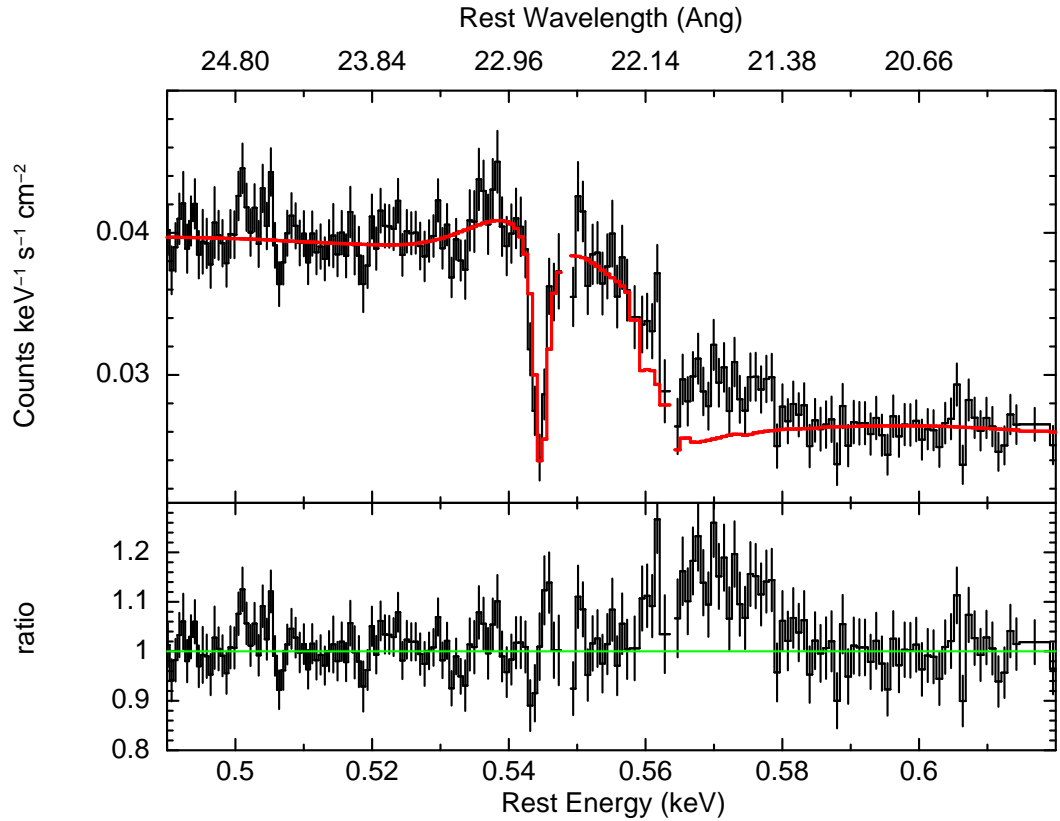}}
\end{center}
\caption{The RGS\,1 count rate spectrum of Ark\,120 in the O K-shell band. The \textsc{tbnew} baseline absorption model as described in Figure\,3 
is shown in red, folded through the instrumental RGS response. Energy (wavelength on the upper-axis) is plotted in the AGN rest frame. A broad O\,\textsc{vii} emission line complex is apparent in the data/model ratio residuals observed by RGS\,1, as observed in the 560-580\,eV region, at energies above the neutral O K-shell absorption due to our Galaxy.} 
\label{Oxygen}
\end{figure}



\clearpage

\begin{figure}
\begin{center}
\rotatebox{-90}{\includegraphics[width=11cm]{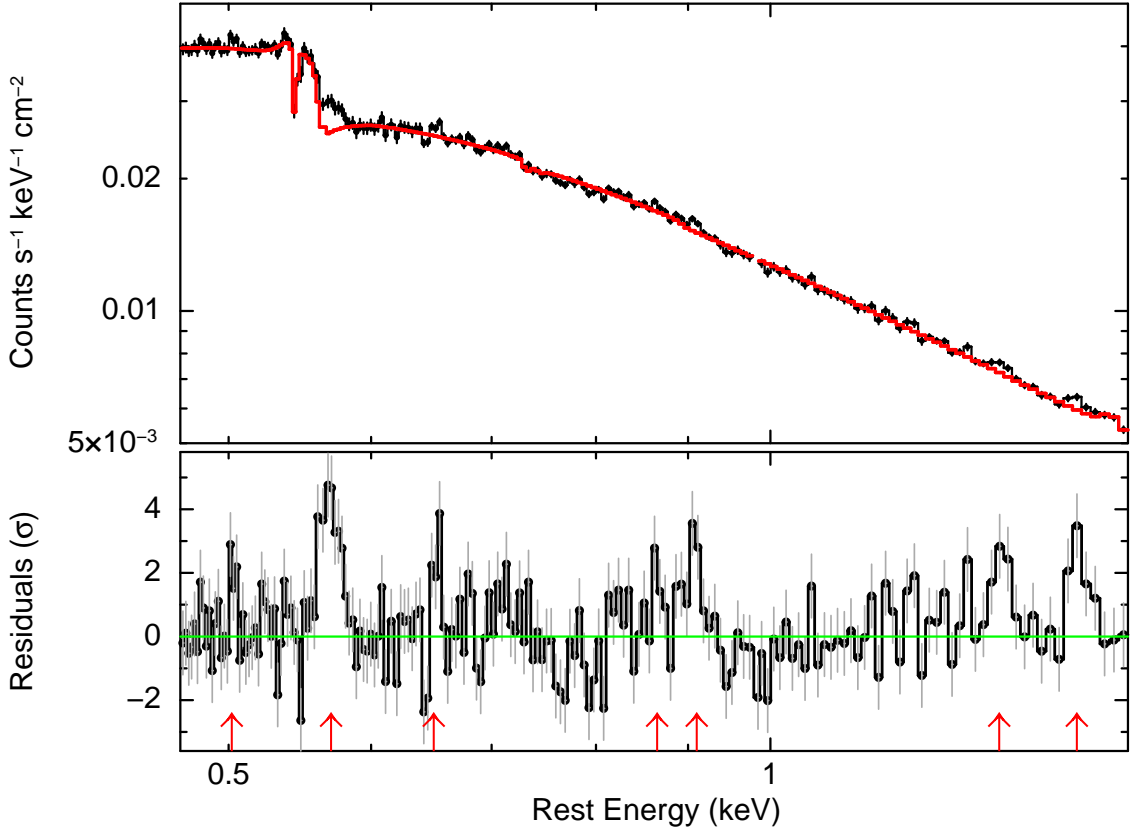}}
\end{center}
\caption{View of the RGS spectrum of Ark 120, plotted over the 0.4-1.6\,keV rest energy band. 
For illustration, the spectrum has been more coarsely binned (compared to the instrumental resolution) to $\Delta\lambda=0.1$\AA\ per bin. The upper panel shows the data points as black points vs rest frame energy (in counts space), while the best fit \textsc{tbnew} ISM absorption 
model (as listed in Table\,2) is overlayed in red. Overall this model provides a good representation of the shape of the X-ray continuum in the RGS band. 
The lower panel shows the residuals (in $\sigma$ units) of the data points compared to this model. Several positive deviations (at $>3\sigma$ confidence) are marked along the x-axis by vertical arrows, which are observed at energies corresponding to 502\,eV, 568\,eV, 654\,eV, 865\,eV, 905\,eV, 1343\,eV and 1476\,eV (to within $\pm1$\,eV uncertainty). These may 
correspond to a series of emission lines, associated with the AGN, due to N\,\textsc{vii} Ly$\alpha$, O\,\textsc{vii} He$\alpha$, O\,\textsc{viii} Ly$\alpha$, Fe\,\textsc{xviii} $3d \rightarrow 2p$, Ne\,\textsc{ix} He$\alpha$, 
Mg\,\textsc{xi} He$\alpha$ and Mg\,\textsc{xii} Ly$\alpha$ respectively. } 
\label{rgsbroad}
\end{figure}

\clearpage

\begin{figure}
\begin{center}
\rotatebox{-90}{\includegraphics[height=10cm]{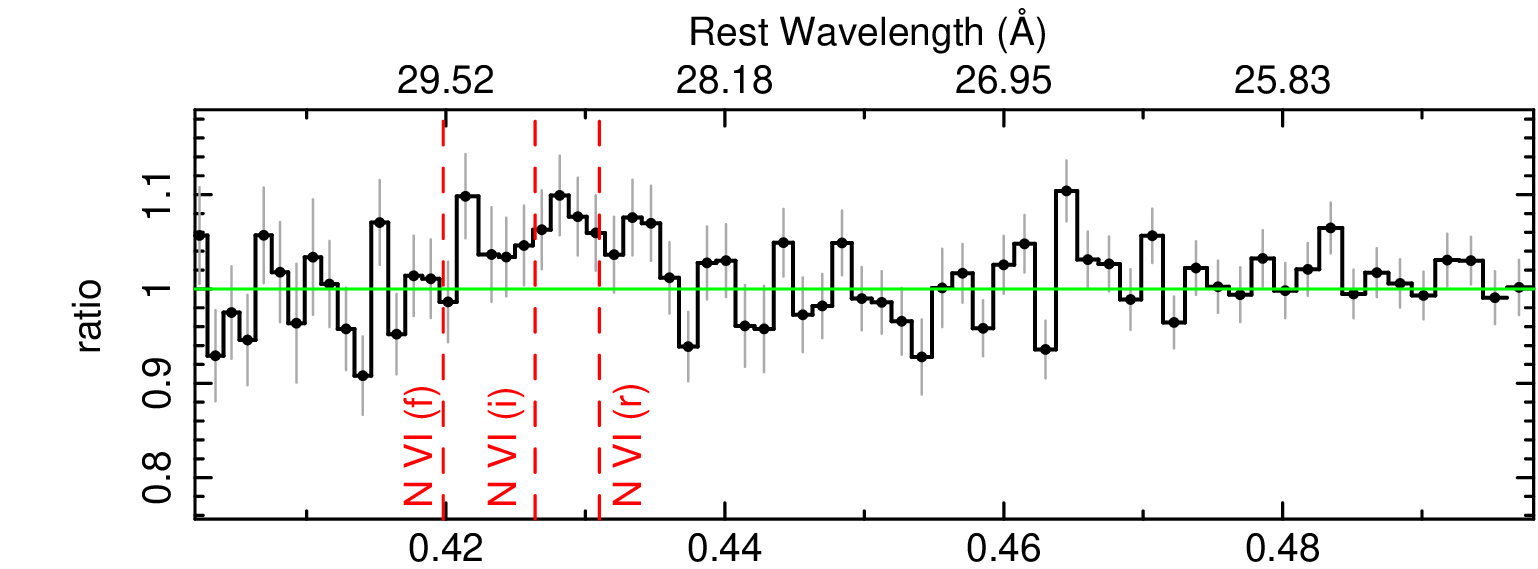}}
\rotatebox{-90}{\includegraphics[height=10cm]{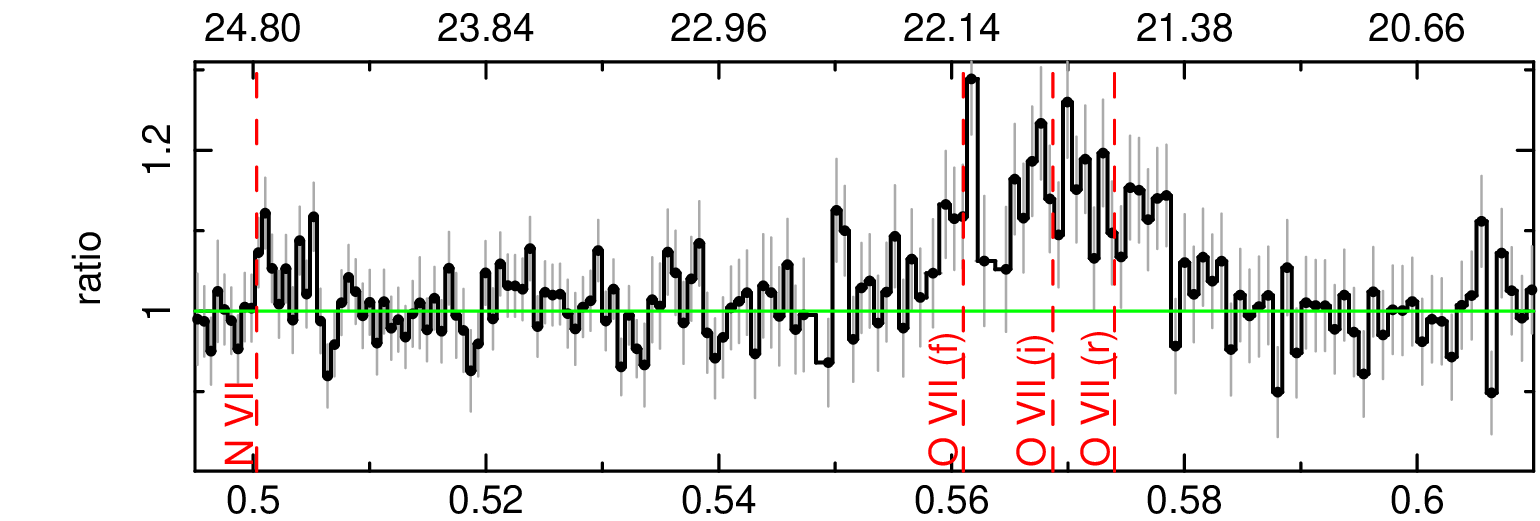}}
\rotatebox{-90}{\includegraphics[height=10cm]{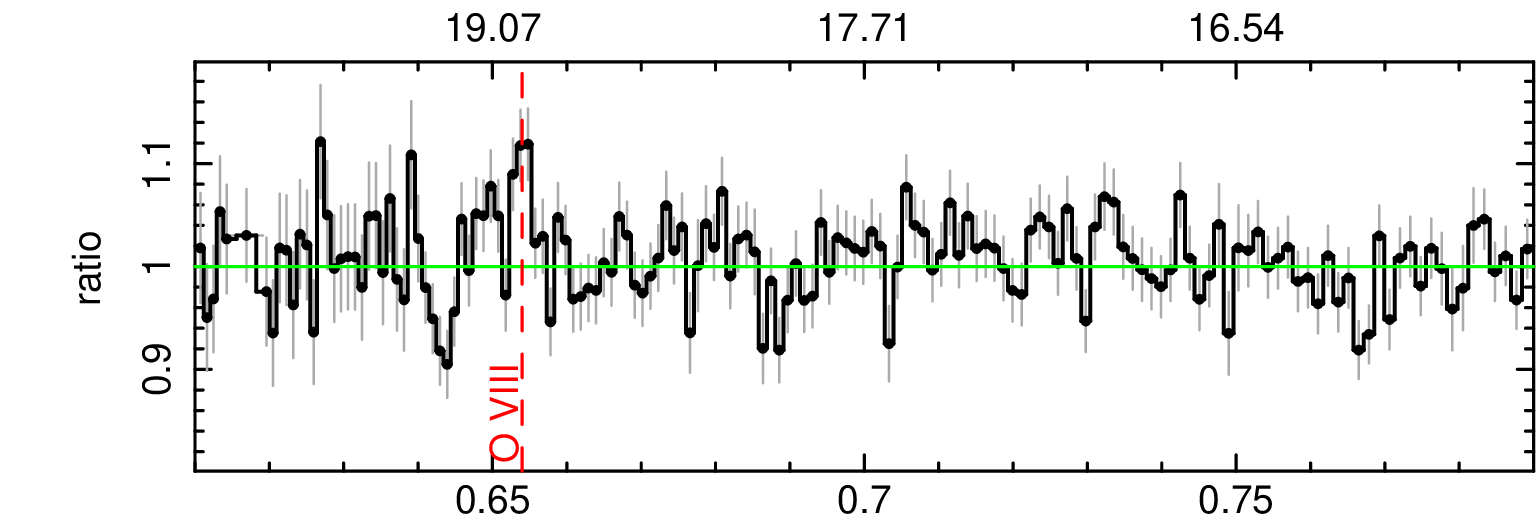}}
\rotatebox{-90}{\includegraphics[height=10cm]{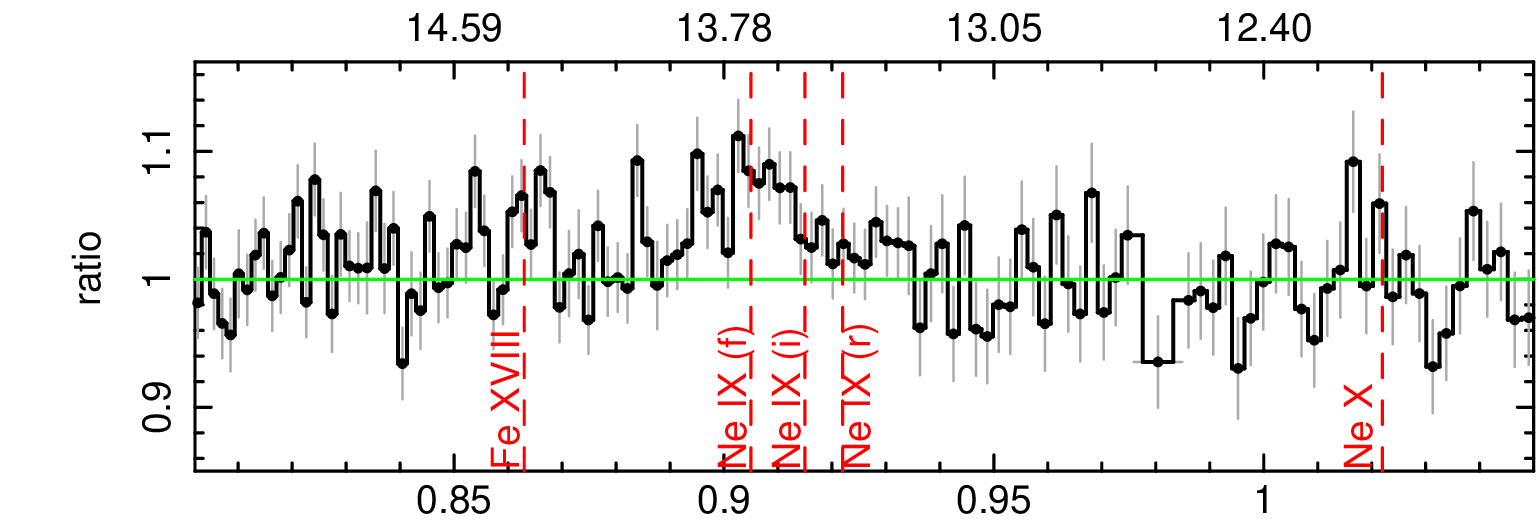}}
\rotatebox{-90}{\includegraphics[height=10cm]{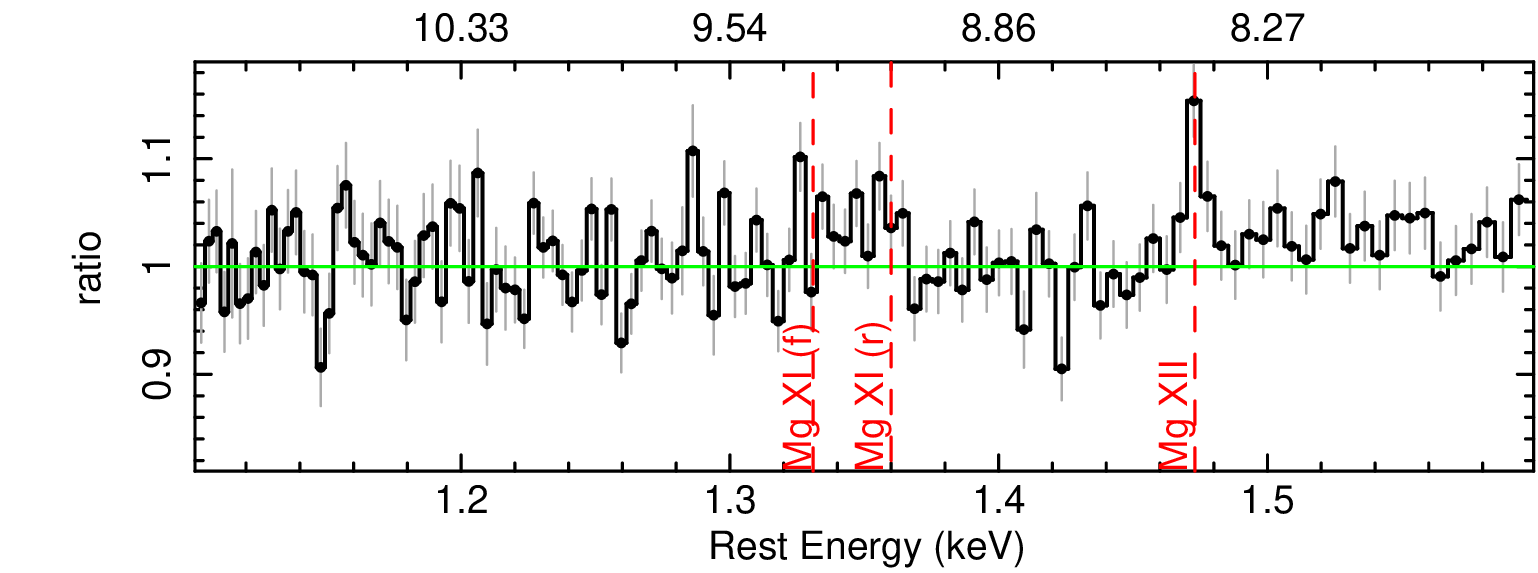}}
\end{center}
\caption{Data/Model ratio residuals of the RGS data to the best-fit ISM absorption model obtained with the baseline \textsc{tbnew} model. Rest energy (in keV) is plotted along the low x-axis, rest wavelength (in Angstroms) along the upper axis. Dashed lines mark the expected positions of the strongest He-like (He$\alpha$) and H-like (Lyman-$\alpha$) emission lines, where the He-like triplets are denoted by the forbidden ($f$), intercombination ($i$) and resonance ($r$) components respectively. The panels (from top to bottom) represent 
the residuals in the region of the N\,\textsc{vi}, O\,\textsc{vii}, O\,\textsc{viii}, Ne\,\textsc{ix}--\textsc{x} and Mg\,\textsc{xi}--\textsc{xii} bands. Note the data are binned at the HWHM resolution of the RGS, except the upper-panel 
around N\,\textsc{vi}, which is binned by a further factor of $\times 3$ for display only.}
\label{rgs-panels2}
\end{figure}

\clearpage

\begin{figure}
\begin{center}
\includegraphics[angle=-90,width=0.42\textwidth]{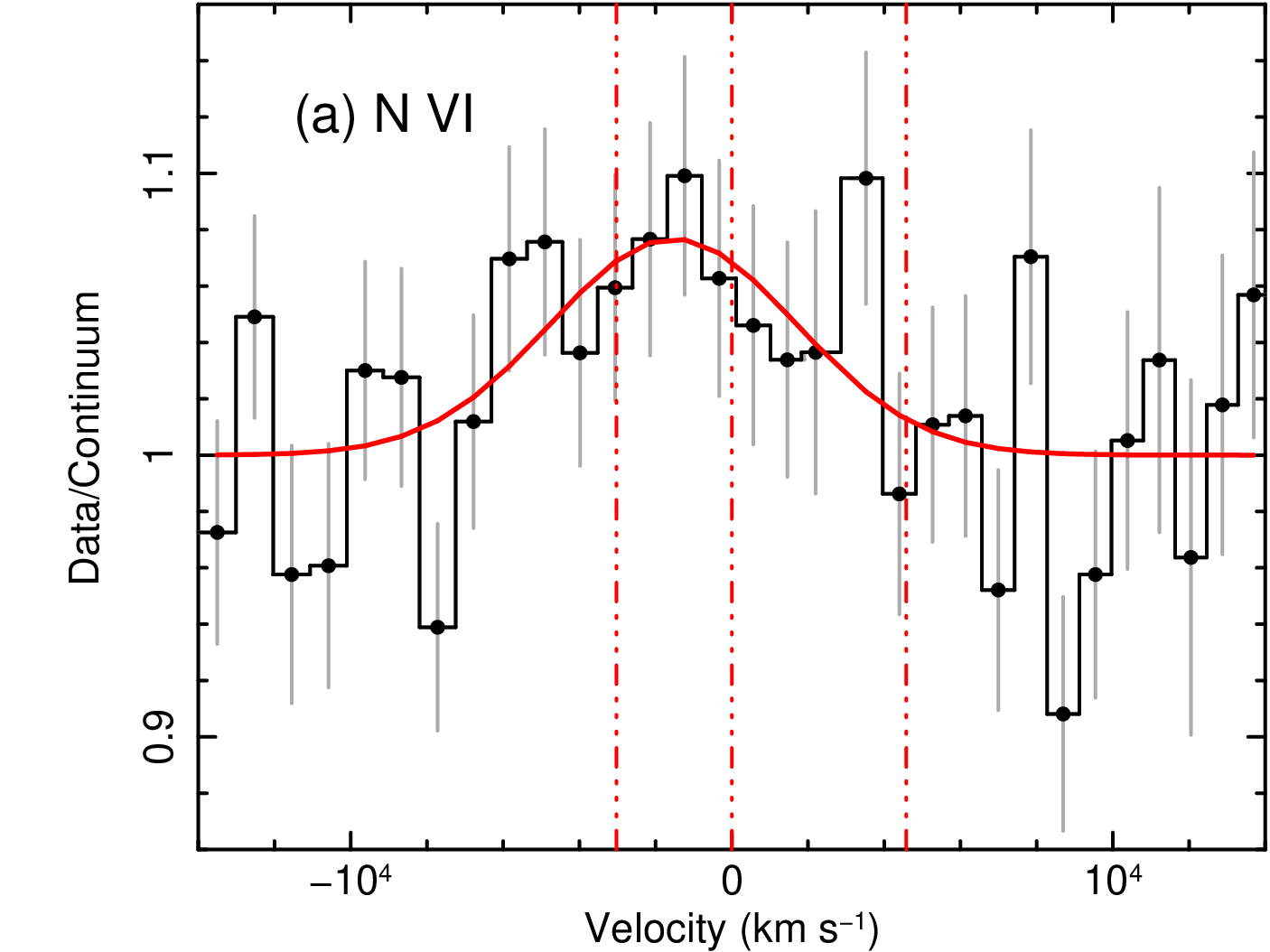}
\includegraphics[angle=-90,width=0.42\textwidth]{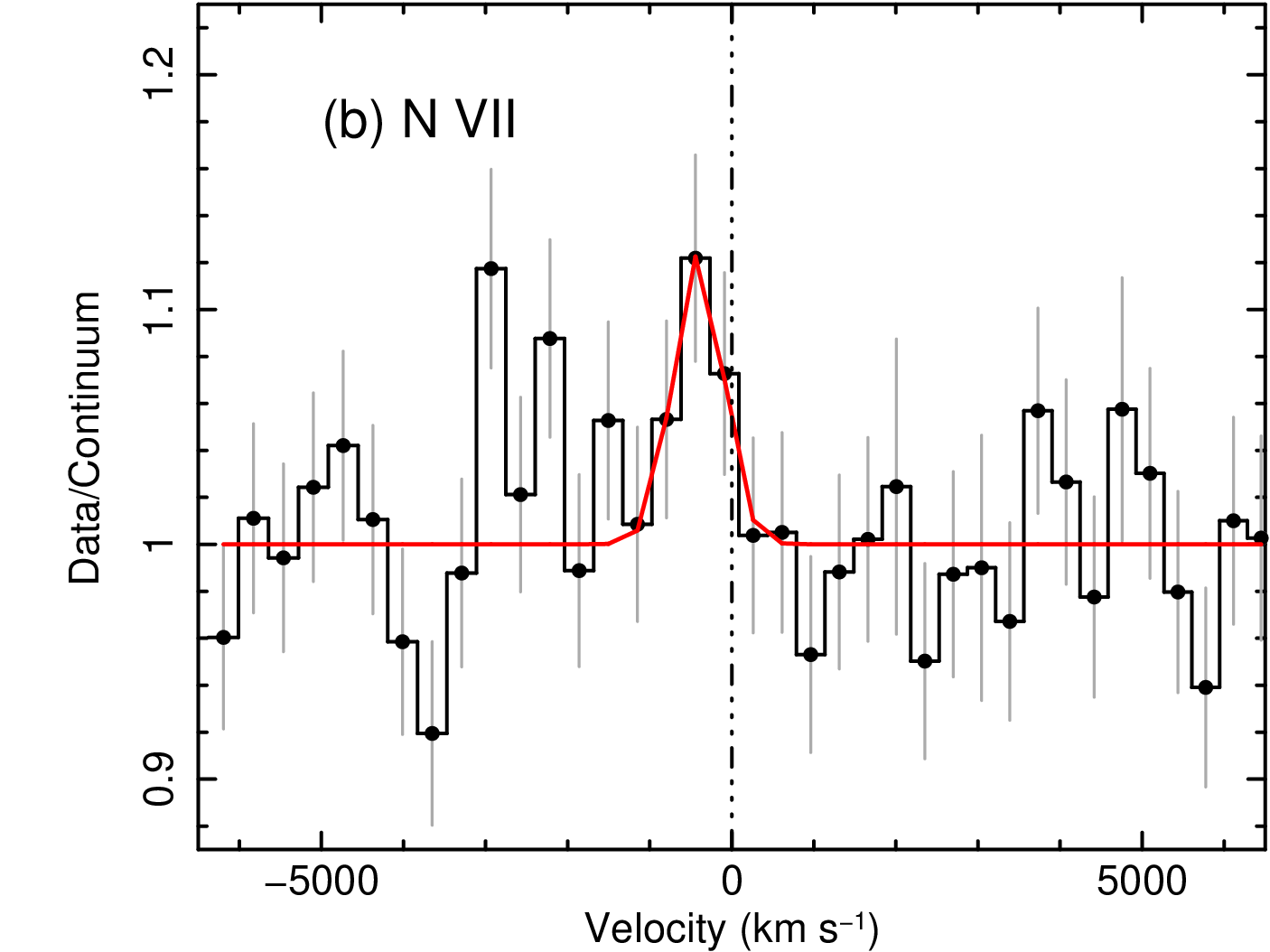}
\includegraphics[angle=-90,width=0.42\textwidth]{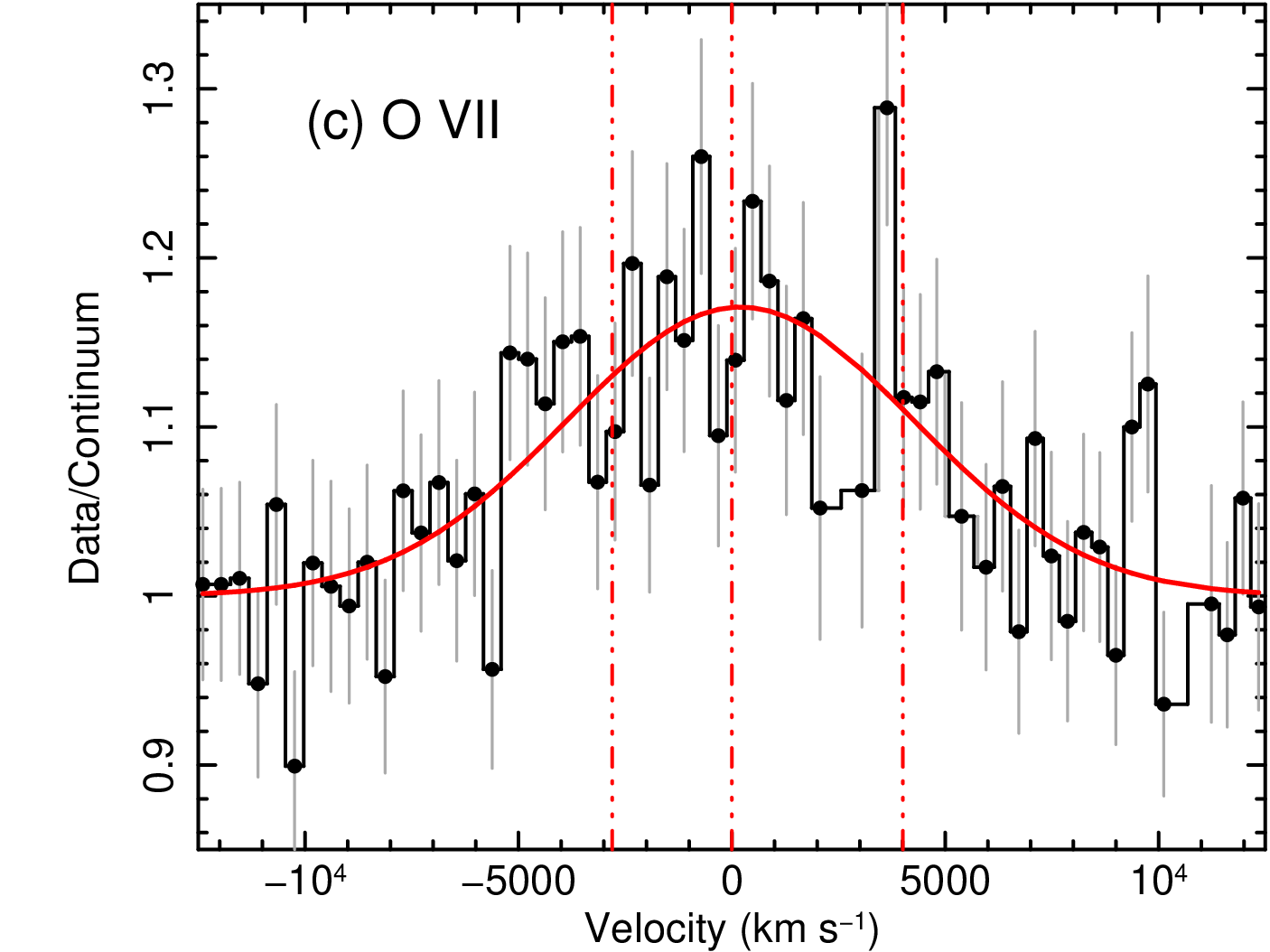}
\includegraphics[angle=-90,width=0.42\textwidth]{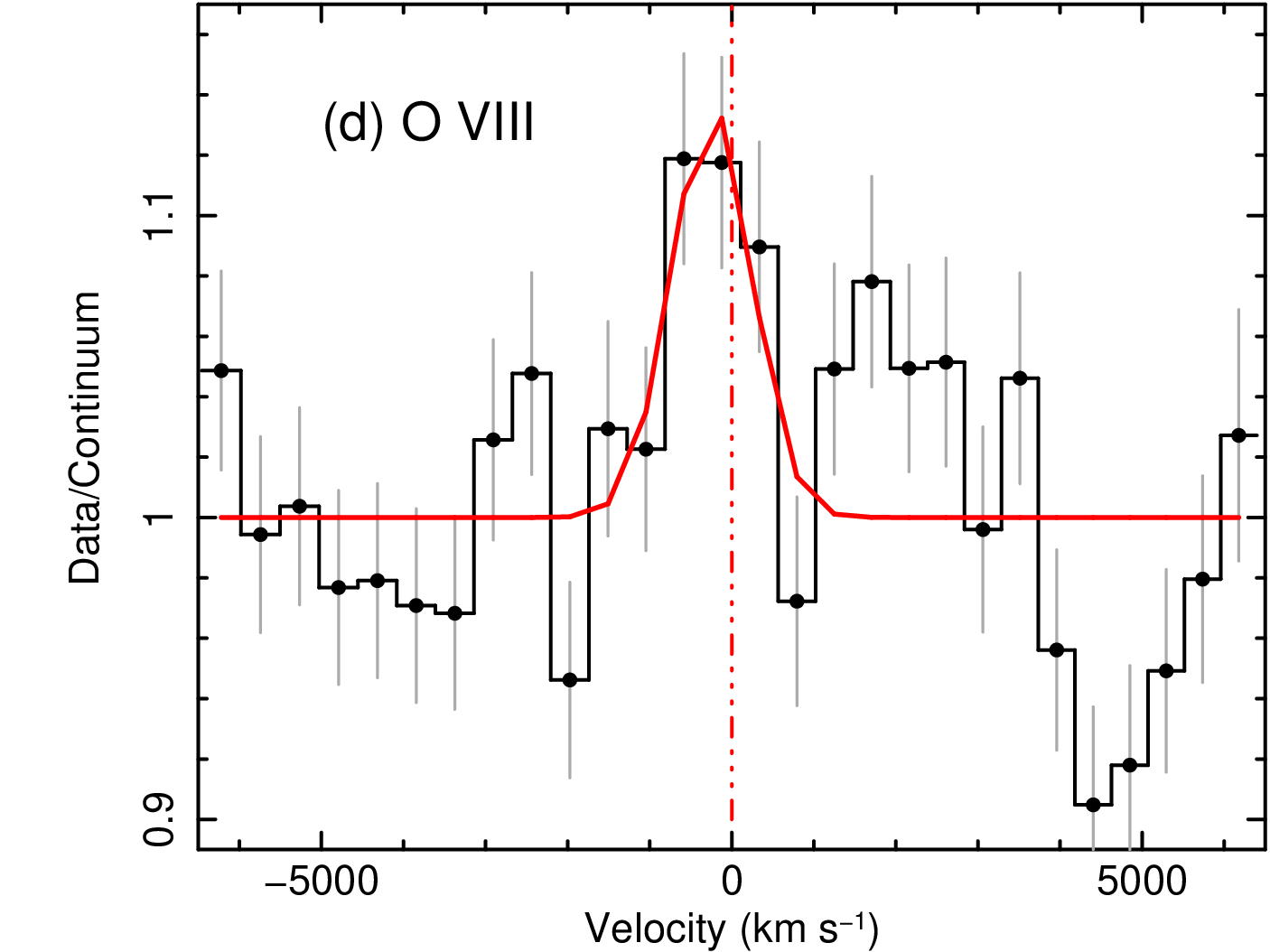}
\includegraphics[angle=-90,width=0.42\textwidth]{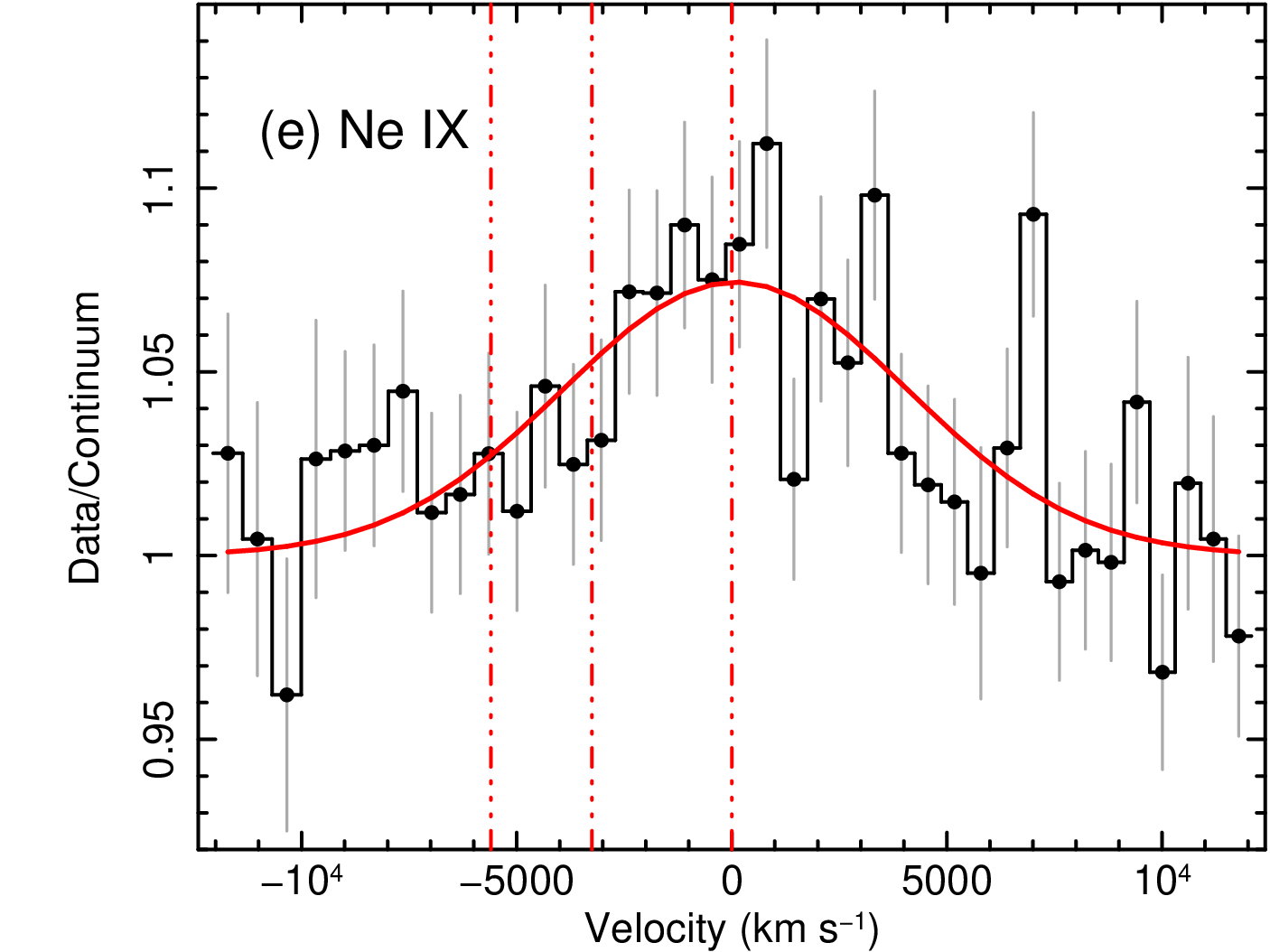}
\includegraphics[angle=-90,width=0.42\textwidth]{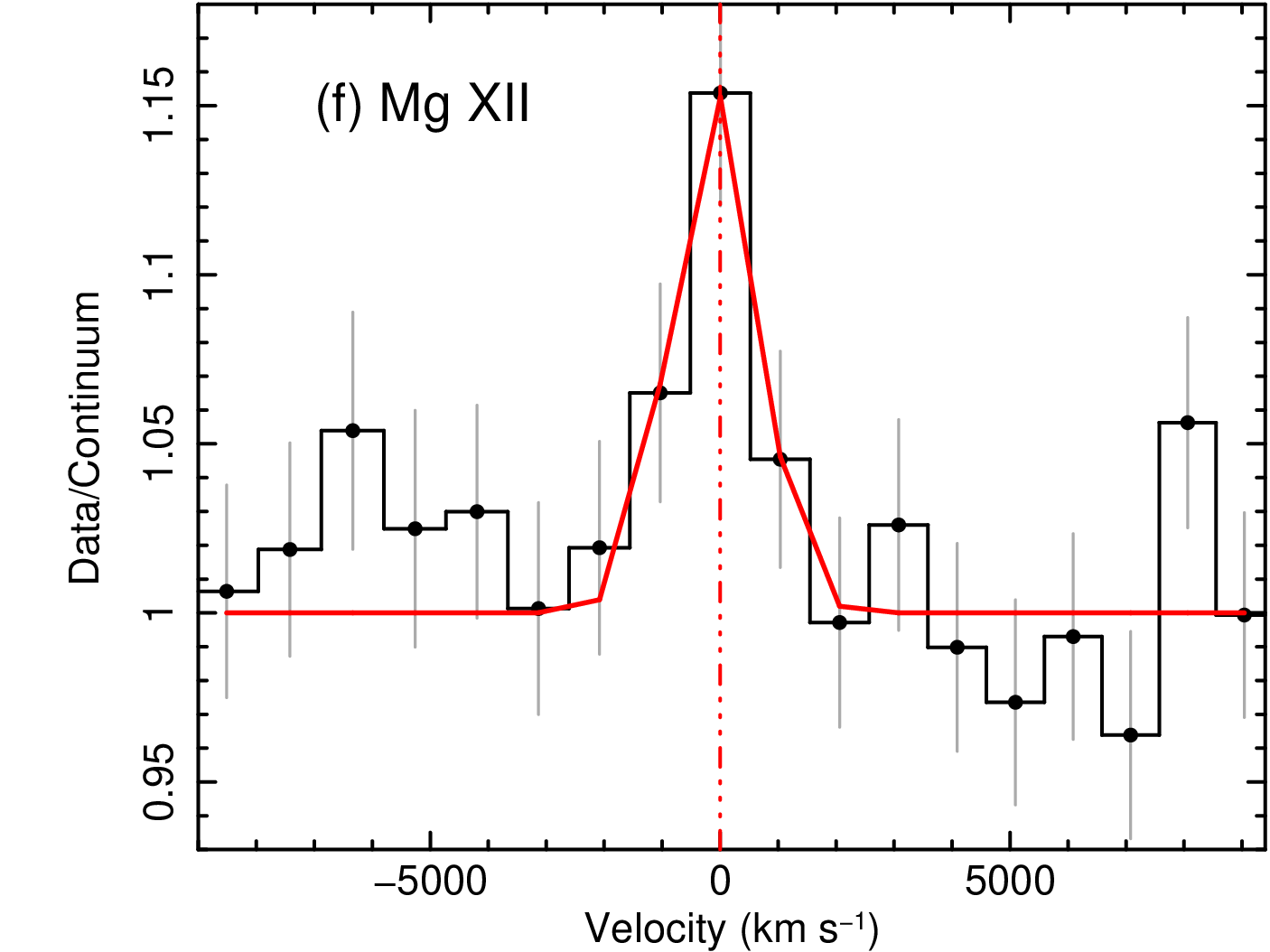}
\end{center}
\caption{Velocity profiles of the main He and H-lines, as measured by XMM-Newton RGS for (a) N\,\textsc{vi}, 
(b) N\,\textsc{vii}, (c) O\,\textsc{vii}, (d) O\,\textsc{viii}, (e) Ne\,\textsc{ix} and (f) Mg\,\textsc{xii}, see Section 5.2 for details. The points (filled circles) show the data divided by the continuum 
model for each line, where negative velocities correspond to blue-shifts. The profiles are centered at zero velocity, 
compared to the known line centroid energies of the doublets for H-like lines, the intercombination emission for 
N\,\textsc{vii} and O\,\textsc{vii} and the forbidden line for Ne\,\textsc{ix}. 
The solid line represents the best-fit single Gaussian emission 
profile for each line profile, with values reported in Table\,5. In the case of the He-like emission (left hand panels), 
the profiles are all resolved and appear broadened, while the expected positions of the resonance, intercombination and forbidden components (left to right) are marked by vertical dot--dashed lines. 
In contrast, the H-like lines (right hand panels) appear unresolved, compared to the instrumental resolution.}
\label{profiles}
\end{figure}

\clearpage

\begin{figure}
\begin{center}
\includegraphics[angle=-90,width=0.42\textwidth]{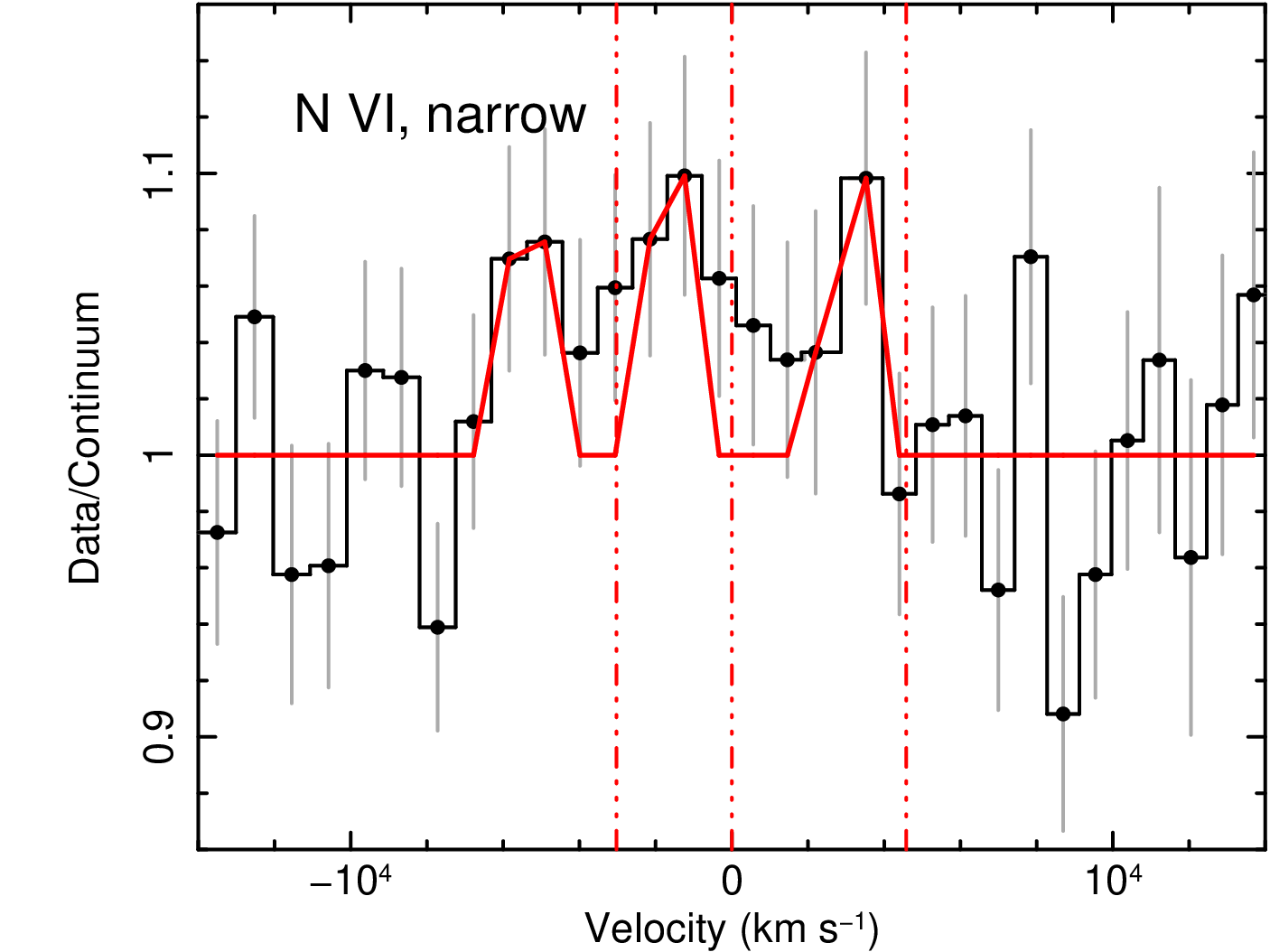}
\includegraphics[angle=-90,width=0.42\textwidth]{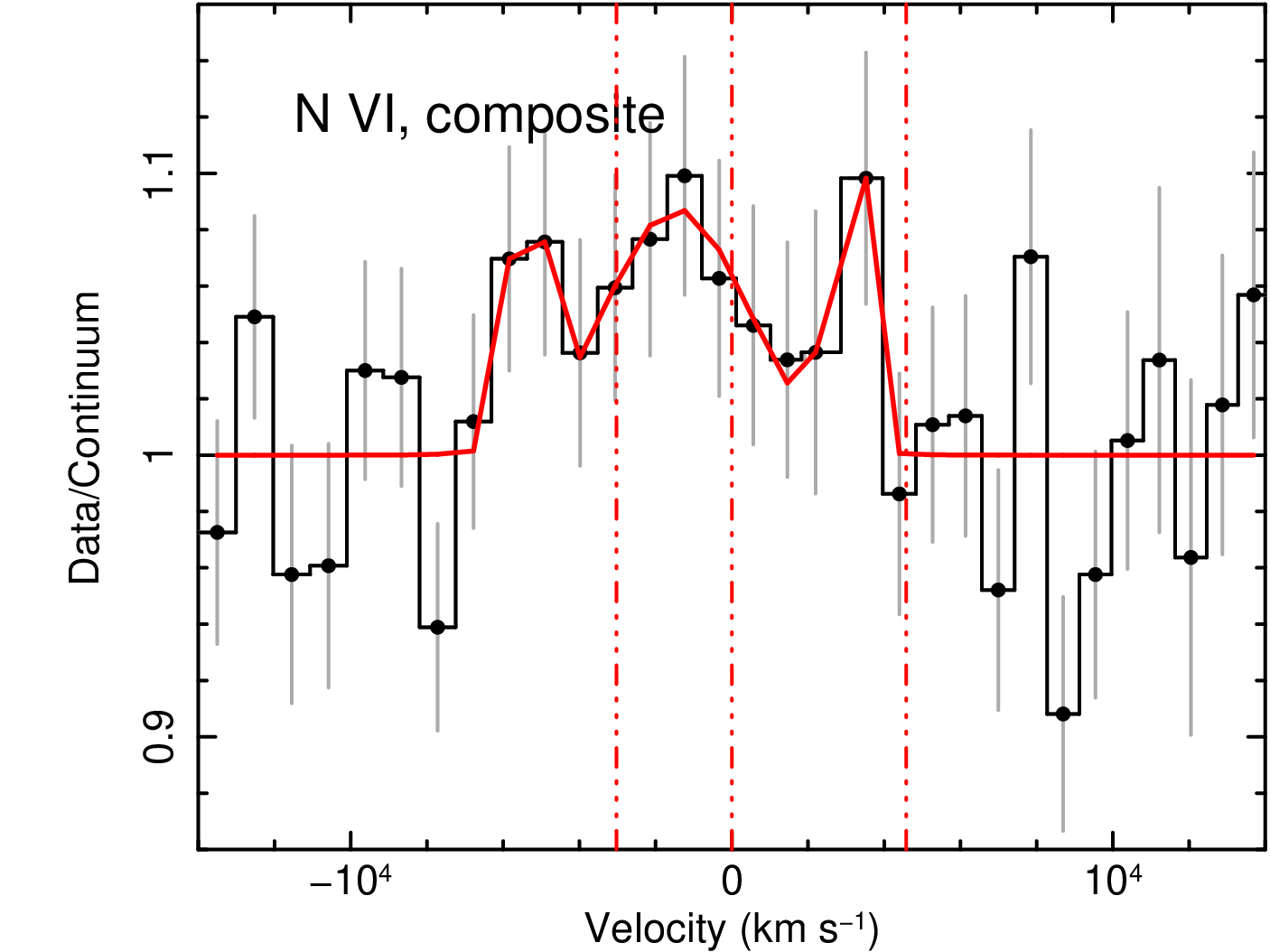}
\includegraphics[angle=-90,width=0.42\textwidth]{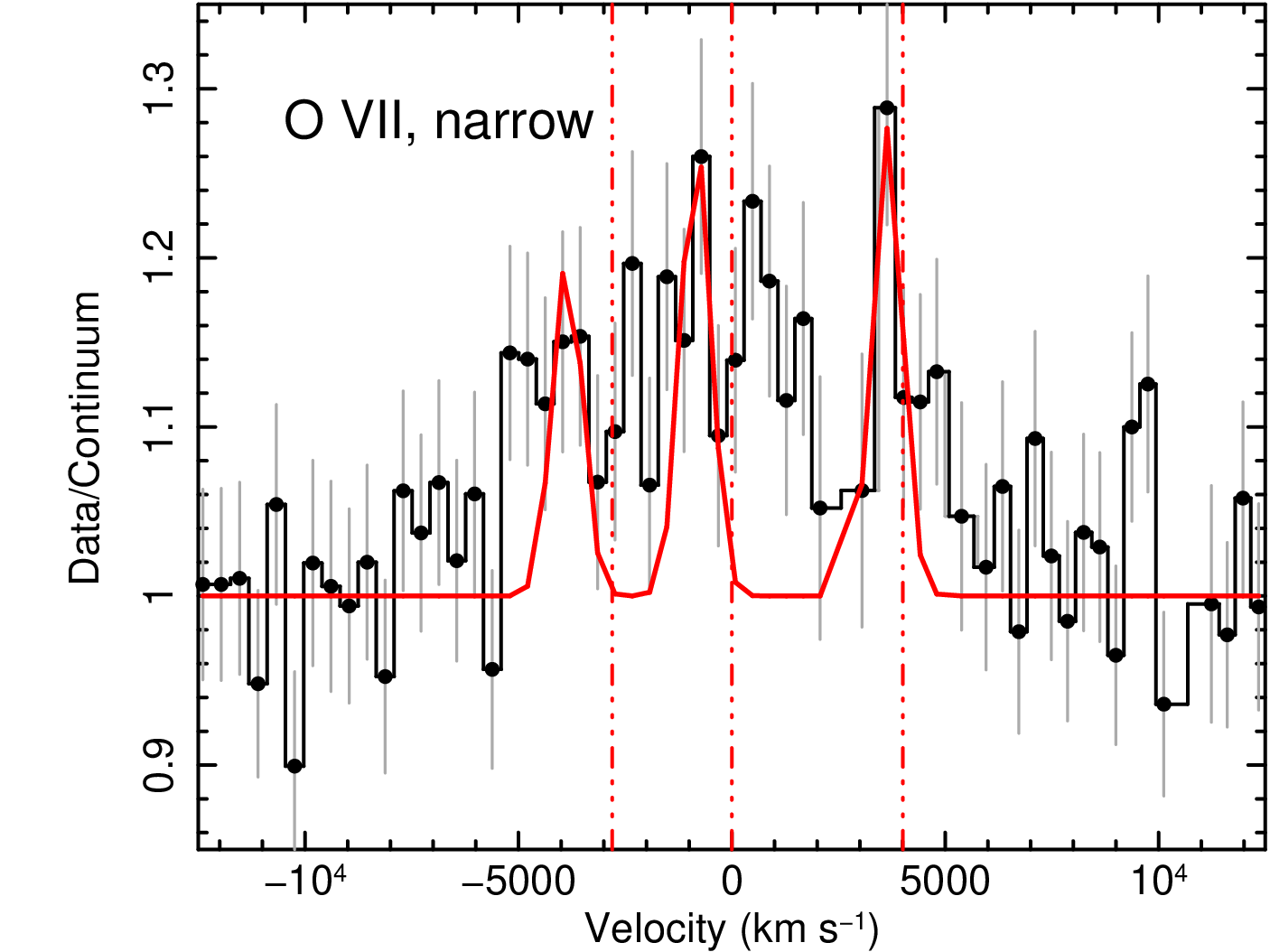}
\includegraphics[angle=-90,width=0.42\textwidth]{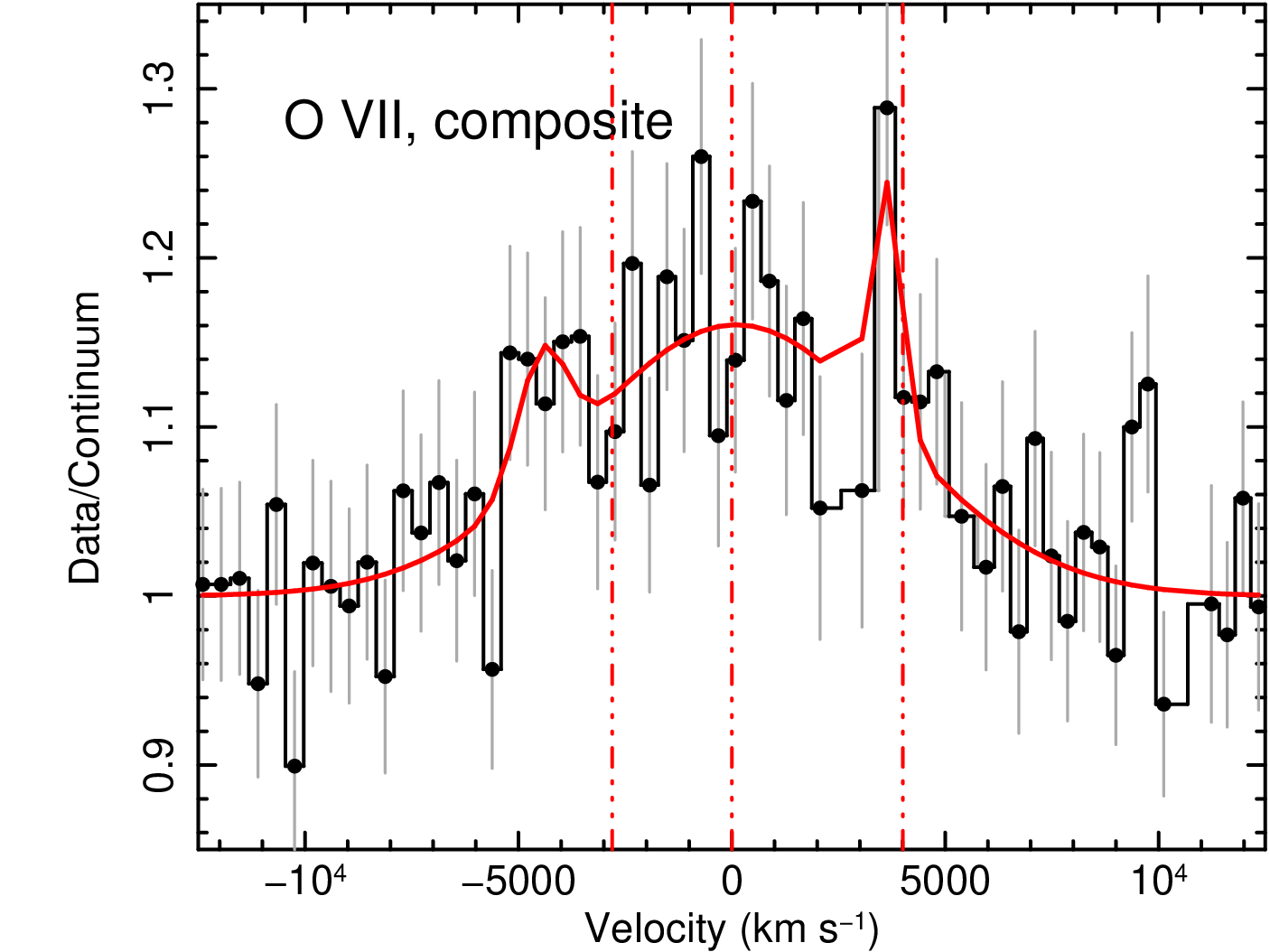}
\includegraphics[angle=-90,width=0.42\textwidth]{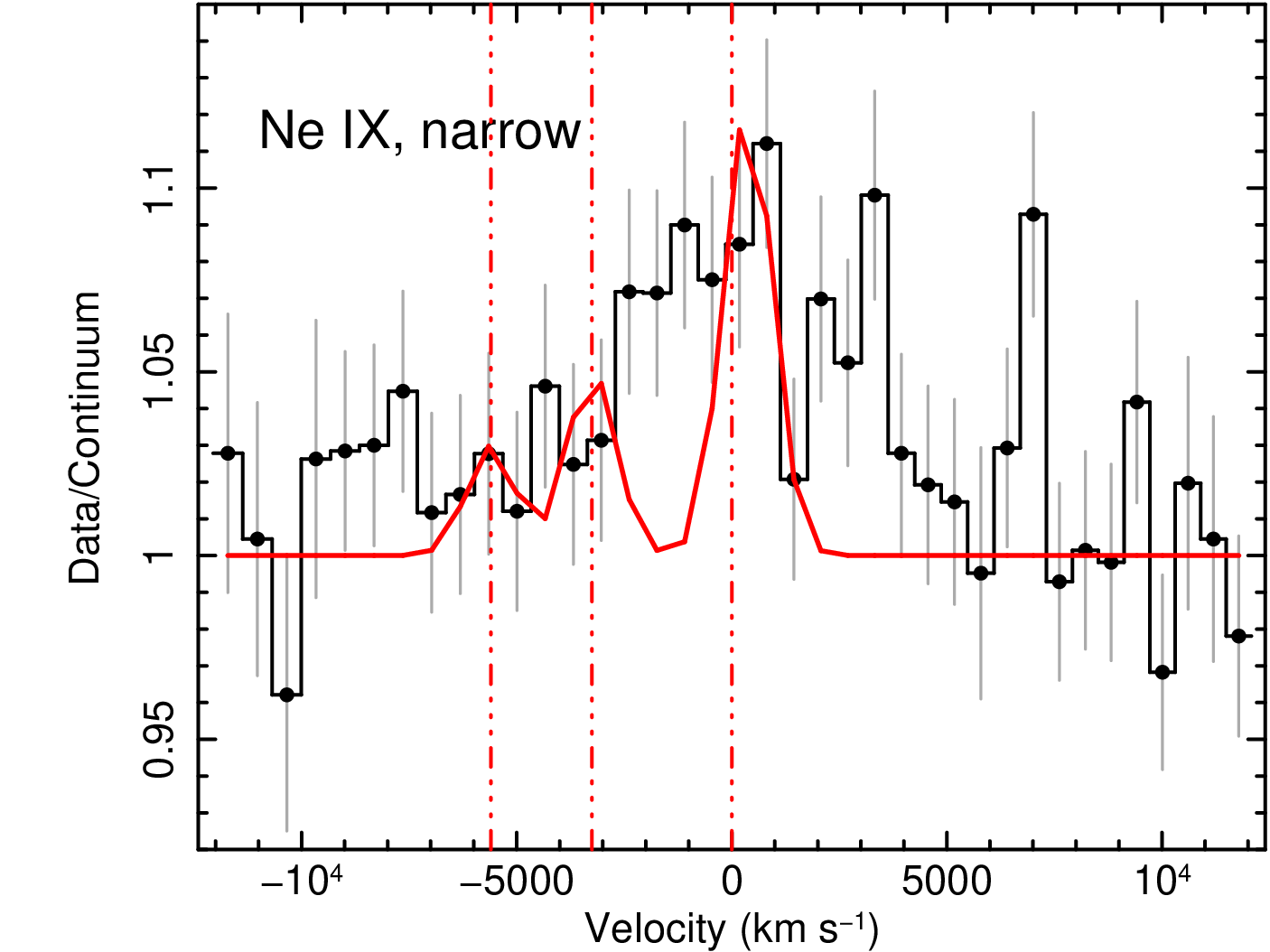}
\includegraphics[angle=-90,width=0.42\textwidth]{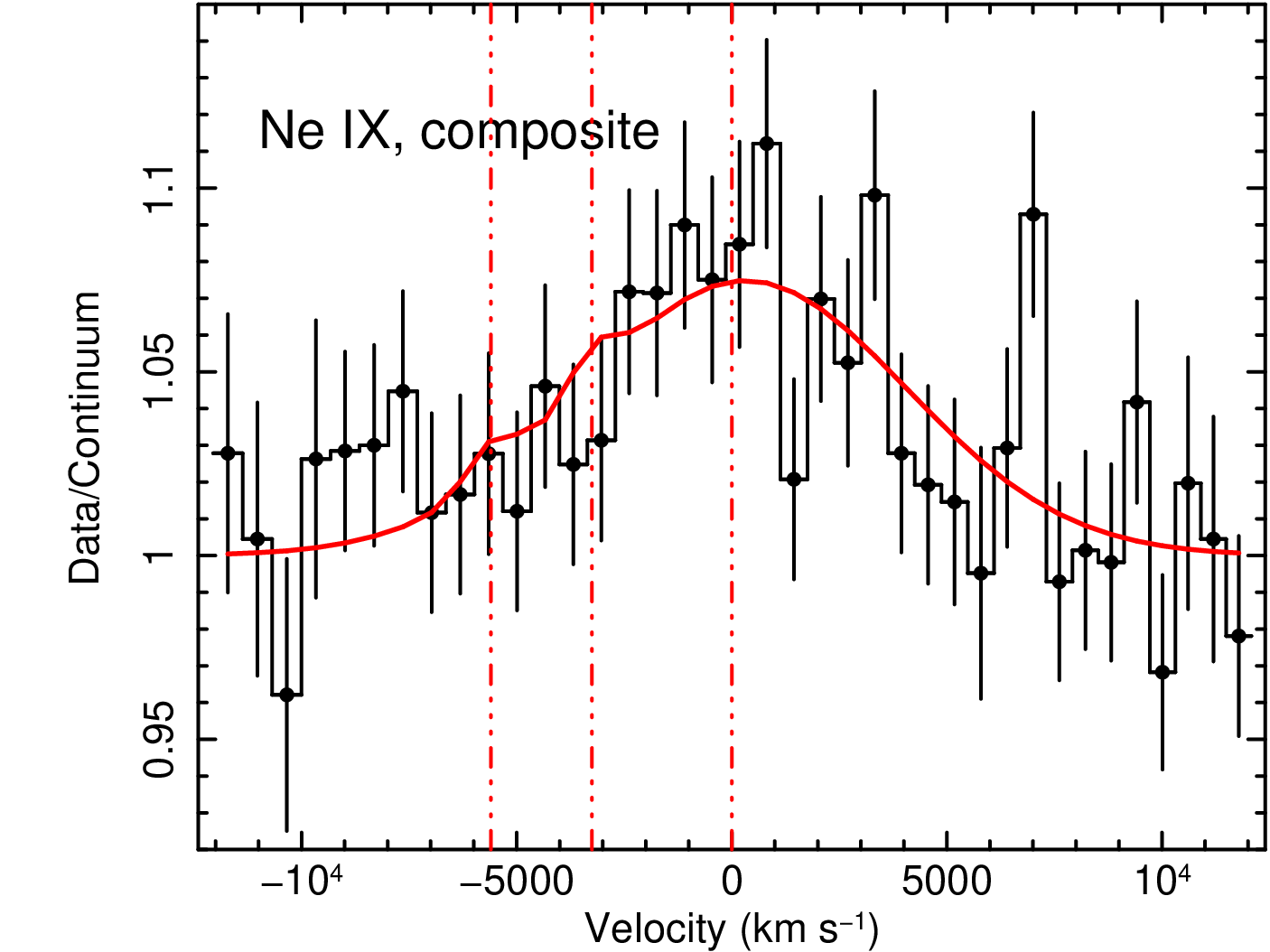}
\end{center}
\caption{As per Figure 7, but showing the fits to the He-like profiles. The left hand panels are the fits to the profiles with a combination of three narrow, unresolved emission lines, corresponding to (from left to right) the resonance, intercombination and forbidden components and allowing for a 
small blue-shift of the lines where required. Overall, the three narrow triplet components provide a 
poor overall representation of the profiles, leaving significant remaining emission unmodeled. 
The right hand panels correspond to composite fits, where a broad emission 
component is included in the line profile, in addition to any narrow line emission where required. For the cases of 
O\,\textsc{vii} and Ne\,\textsc{ix} in particular, the 
profiles are dominated by an underlying broad (FWHM $\sim 8000$\,km\,s$^{-1}$) line, corresponding to intercombination emission for O\,\textsc{vii} and forbidden emission for Ne\,\textsc{ix}, while any contributions from narrow components are small. 
The profiles indicate that most of the He-like line emission arises from denser gas, of the order $\sim 10^{11}$\,cm$^{-3}$, with velocities commensurate with the AGN Broad Line Region.}
\label{Helike}
\end{figure}

\clearpage

\begin{figure}
\begin{center}
\rotatebox{-90}{\includegraphics[width=11cm]{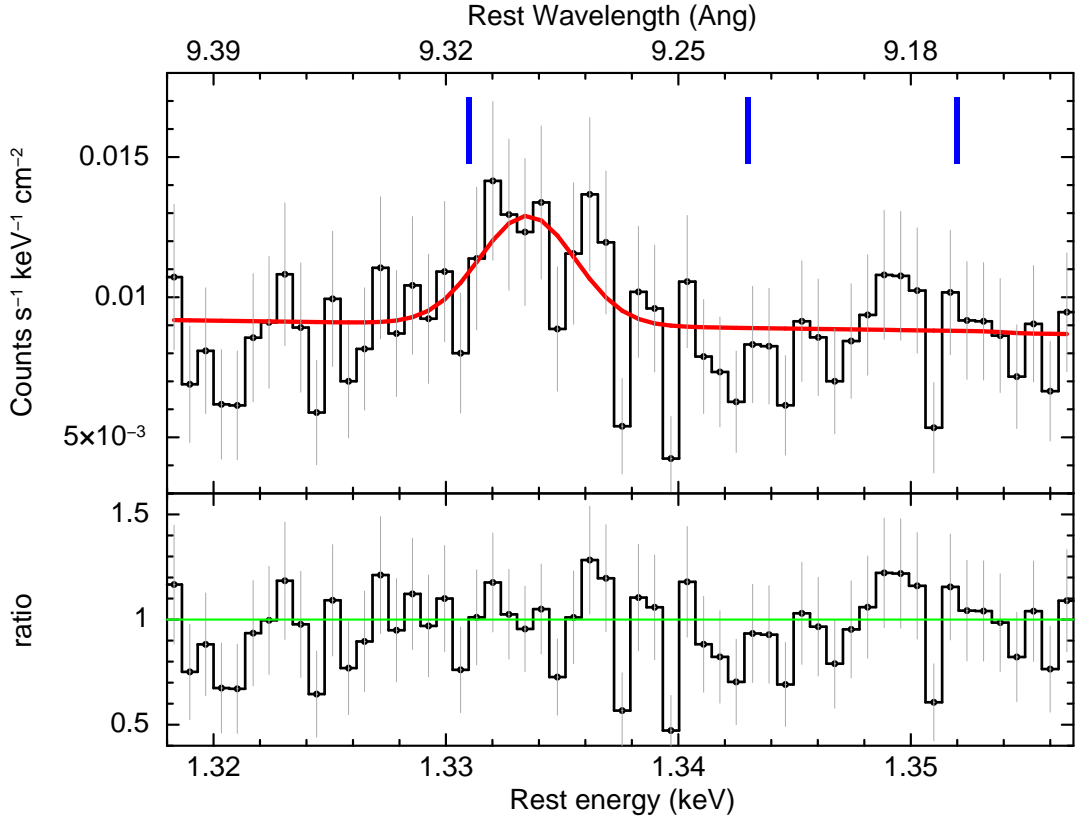}}
\end{center}
\caption{Portion of the Chandra HEG spectrum of Ark\,120, in the region of the Mg\,\textsc{xi} He-like triplet, 
plotted in the AGN rest frame. The expected positions (from left to right) of the forbidden, intercombination
and resonance line components are marked with blue vertical lines. Significant line emission is detected 
just bluewards of the expected position of the forbidden line, while no emission is required from either of the 
intercombination or resonance components. Due to the high resolution of the HEG spectrum, the 
forbidden line is resolved, with a width of 
$\sigma_{v}=450^{+290}_{-180}$\,km\,s$^{-1}$ and likely corresponds to the narrow component of 
emission which is unresolved in the RGS. Note that the spectral bins sample the HWHM resolution of the HEG, 
which corresponds to $\Delta\lambda=0.005$\AA\ or $\sim160$\,km\,s$^{-1}$ at the position of the Mg\,\textsc{xi} triplet.}  
\label{Mgprofile}
\end{figure}

\clearpage

\begin{figure}
\begin{center}
\rotatebox{-90}{\includegraphics[width=11cm]{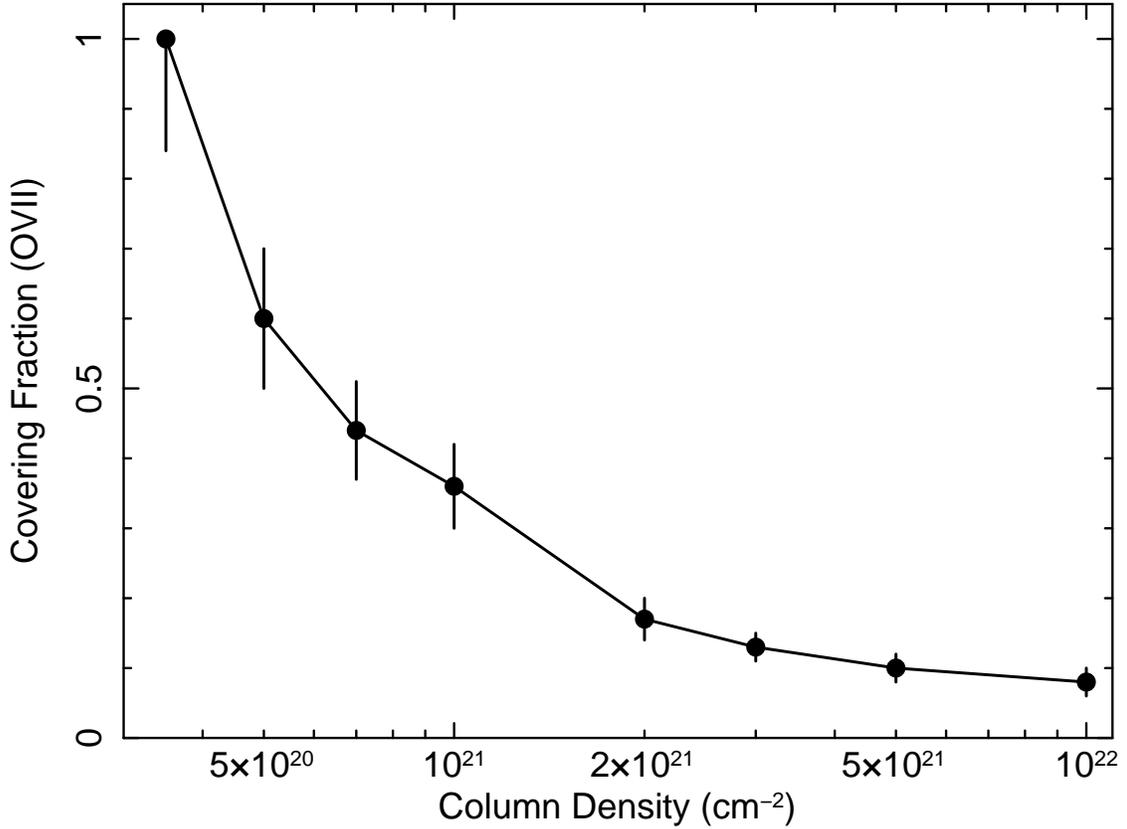}}
\end{center}
\caption{Geometric covering fraction (as a fraction of $4\pi$ steradian) 
of the broad line soft X-ray emitting gas responsible for the broad O\,\textsc{vii} profile. This has been fitted 
as a function of the emitter column density, from the photoionized 
emission model discussed in Section\,6.4. The covering fraction for the O\,\textsc{vii} gas decreases with column density, with a minimum covering fraction of 
$f=0.1$ (i.e. 10\% of $4\pi$\,steradian) deduced for columns of 
$N_{\rm H}=10^{22}$\,cm$^{-2}$ and higher.  
A minimum column density of $N_{\rm H}>3\times10^{20}$\,cm$^{-2}$ is required 
corresponding to where the gas 
fully covers the AGN with $f=1$. 
Overall the deduced parameters appear consistent with the expected gas 
distribution in the AGN Broad Line Region.}  
\label{covering}
\end{figure}

\clearpage

\begin{deluxetable}{lccccc}
\tabletypesize{\small}
\tablecaption{Summary of Ark\,120 Observations}
\tablewidth{0pt}
\tablehead{
\colhead{Mission} & \colhead{Obsid} & \colhead{Start Date/Time$^{a}$} & \colhead{Inst} 
& \colhead{Exposure (ks)} & \colhead{Net Rate s$^{-1}$}}

\startdata
  XMM-Newton & 0721600201 & 2014-03-18 08:52:49 & RGS\,1 & 108.8 & $0.795\pm0.003$ \\
& -- & -- & RGS\,2 & -- & $0.909\pm0.003$ \\
& 0721600301 & 2014-03-20 08:58:47 & RGS\,1 & 116.6 & $0.653\pm0.003$ \\
& -- & -- & RGS\,2 & -- & $0.747\pm0.003$ \\
& 0721600401 & 2014-03-22 08:25:17 & RGS\,1 & 102.8 & $0.718\pm0.003$ \\
& -- & -- & RGS\,2 & -- & $0.831\pm0.003$\\
& 0721600501 & 2014-03-24 08:17:19 & RGS\,1 & 103.6 & $0.654\pm0.003$ \\
& -- & -- & RGS\,2 & -- & $0.746\pm0.003$\\
& Total & -- & RGS\,1 & 431.9$^{b}$ & $0.705\pm0.001$ \\
&  & -- & RGS\,2 & 430.8$^{b}$ & $0.808\pm0.001$ \\
&  & -- & RGS\,1+2 & 862.7$^{b}$ & $0.756\pm0.001$ \\
\hline
Chandra & 16539 & 2014-03-17 07:47:58 & MEG & 63.0 & $0.914\pm0.004$ \\
        & -- & -- & HEG & -- & $0.417\pm0.003$ \\        
& 15636 & 2014-03-21 04:38:43 & MEG & 10.2 & $0.752\pm0.011$ \\
& -- & -- & HEG & -- & $0.356\pm0.009$ \\
& 16540 & 2014-03-22 13:55:12 & MEG & 47.3 & $0.832\pm0.005$ \\
& -- & -- & HEG & -- & $0.389\pm0.004$ \\
& Total & -- & MEG & 120.5 & $0.868\pm0.003$ \\
& -- & -- & HEG & -- & $0.401\pm0.002$ \\
\enddata

\tablenotetext{a}{Observation Start/End times are in UT.} 
\tablenotetext{b}{Net exposure time, after screening and deadtime correction, in ks.}
\label{observations}
\end{deluxetable}

\clearpage

\begin{deluxetable}{llc}
\tabletypesize{\small}
\tablecaption{Baseline Model Parameters to 2014 Ark\,120 RGS Spectrum}
\tablewidth{0pt}
\tablehead{
\colhead{Component} & \colhead{Parameter} & \colhead{Value}}
\startdata
Continuum:-\\
Galactic absorption & $N_{\rm H}$$^{a}$ & $8.8\pm0.4$ \\
& $A_{O}$$^{b}$ & $1.62\pm0.10$ \\
Powerlaw & $\Gamma$ & $2.06\pm0.06$ \\
& $N_{\rm PL}$$^{c}$ & $1.38\pm0.04$ \\ 
Blackbody & $kT^{d}$ & $126\pm 6$ \\
& $N_{\rm BB}^{e}$ & $9.7\pm1.7$ \\\hline
Absorption Lines:-\\
N\,\textsc{i} K$\alpha$$^g$ & $E_{\rm obs}$ ($\lambda_{\rm obs}$) & $396.6\pm0.5$ [31.26] \\
& $N_{\rm ph}$ & $-7.2\pm0.7$ \\
& EW & $0.59\pm0.06$ \\
O\,\textsc{i} K$\alpha$$^g$ & $E_{\rm obs}$ ($\lambda_{\rm obs}$) & $527.2\pm0.2$ [23.52] \\
& $N_{\rm ph}$ & $-7.0\pm1.0$ \\
& EW & $1.14\pm0.16$ \\\hline
Flux & $F_{0.4-2.0}$$^{h}$ & 2.95 \\
\enddata
\tablenotetext{a}{Galactic neutral absorption column, modeled with tbnew, units $\times10^{20}$\,cm$^{-2}$.}
\tablenotetext{b}{O abundance, compared to the Solar abundance table of Wilms et al. (2000).}
\tablenotetext{c}{Power law photon flux at 1 keV, in units of $\times 10^{-2}$\,photons\,cm$^{-2}$\,s$^{-1}$\,keV$^{-1}$.}
\tablenotetext{d}{Blackbody temperature, units eV.}
\tablenotetext{e}{Blackbody normalization, in units of $\times 10^{-5} \frac{L_{39}}{D_{\rm 10 kpc}^2}$, 
where $L_{39}$ is the luminosity in units of $\times 10^{39}$\,erg\,s$^{-1}$ and $D_{\rm 10 kpc}$ 
is the source distance in units of 10\,kpc.} 
\tablenotetext{g}{N\,\textsc{i} and O\,\textsc{i} K$\alpha$ absorption line parameters. Energy ($E_{\rm obs}$) (or wavelength $\lambda_{\rm obs}$ in brackets) are 
given in the observed ($z=0$) frame in units of eV (or \AA); photon flux ($N_{\rm ph}$) is in units of $\times10^{-5}$\, 
photons\,cm$^{-2}$\,s$^{-1}$; equivalent width (EW) is in units of eV.} 
\tablenotetext{h}{Observed (absorbed) continuum flux from 0.4--2.0\,keV in units of $\times 10^{-11}$\,erg\,cm$^{-2}$\,s$^{-1}$.}
\label{parameters}
\end{deluxetable}

\clearpage

\begin{deluxetable}{lccccc}
\tabletypesize{\small}
\tablecaption{Galactic ISM Absorption towards Ark\,120.}
\tablewidth{0pt}
\tablehead{
\colhead{Parameter} & \colhead{\textsc{tbvarabs}$^a$} & \colhead{\textsc{tbnew}$^{a}$} & \colhead{\textsc{ismabs}$^a$}
&  \colhead{\textsc{tbnew}$^2$$^{a}$} & \colhead{\textsc{ismabs}$^{2}$$^a$}}
\startdata
$N_{\rm H}$$^{b}$ & $9.4\pm0.4$ & $8.8\pm0.4$ & $8.51\pm0.4$ & $11.1\pm0.4$ & $11.4\pm0.6$ \\
$N_{\rm O}$$^{c}$ & $6.4\pm0.5$ & $7.0\pm0.5$ & $6.5\pm0.3$ & $6.5\pm0.3$ & $6.1\pm0.3$ \\
${\rm O}/{\rm H}$$^{d}$ & $6.8\pm0.5$ & $7.9\pm0.5$ & $7.7\pm0.5$ & $5.9\pm0.3$ & $5.4\pm0.4$ \\
$A_{\rm O}$$^{e}$ & $1.39\pm0.10$ & $1.62\pm0.10$ & $1.57\pm0.10$ & $1.20\pm0.04$ & $1.10\pm0.08$ \\
\hline
$N_{\rm N}$$^{f}$ & -- & -- & $0.72\pm0.27$ & -- & $0.76\pm0.25$ \\
${\rm N}/{\rm H}$$^{g}$ & -- & -- & $8.5\pm3.2$ & -- & $6.7\pm2.2$ \\
\hline
$N_{\rm Ne}$$^{f}$ & -- & -- & $1.2\pm0.3$ & -- & $1.4\pm0.3$ \\
${\rm Ne}/{\rm H}$$^{g}$ & -- & -- & $14\pm4$ & -- & $12\pm3$ \\
\hline
$N_{\rm Fe}$$^{f}$ & -- & -- & $0.26\pm0.06$ & -- & $0.25\pm0.05$ \\
${\rm Fe}/{\rm H}$$^{g}$ & -- & -- & $3.0\pm0.7$ & -- & $2.2\pm0.5$ \\
\hline
$N_{\rm O^+}$$^{h}$ & -- & -- & $<0.42$ & -- & --\\
$N_{\rm O^{2+}}$$^{h}$ & -- & -- & $<0.06$ & -- & -- \\
\hline
$\chi^{2}/{\rm dof}$ & $1392.7/934$ & $1087.3/914$ & $1086.6/911$ & $1110.8/914$ & $1093.6/911$
\enddata
\tablenotetext{a}{Galactic ISM absorption model, fitted with either \textsc{tbvarabs}, \textsc{tbnew} or \textsc{ismabs}. 
In the first three cases, the continuum is fitted with a power-law plus blackbody model to the RGS band, 
while in the latter \textsc{tbnew}$^2$ and \textsc{ismabs}$^{2}$ cases, the continuum is fitted with a power-law plus Comptonized blackbody spectrum.}
\tablenotetext{b}{Galactic neutral hydrogen column, in units $\times10^{20}$\,cm$^{-2}$.}
\tablenotetext{c}{Neutral Oxygen column, units $\times10^{17}$\,cm$^{-2}$.}
\tablenotetext{d}{Derived ratio of O to H column density, $\times10^{-4}$.}
\tablenotetext{e}{Relative O/H abundance, compared to the Solar abundance value of  $4.9\times10^{-4}$ in 
 Wilms et al. (2000).}
\tablenotetext{f}{Column density of neutral N, Ne or Fe, in units $\times10^{17}$\,cm$^{-2}$.}
\tablenotetext{g}{Derived ratio of ${\rm N}/{\rm H}$, ${\rm Ne}/{\rm H}$ and ${\rm Fe}/{\rm H}$ column densities, $\times10^{-5}$. Note the relative Solar abundances listed in Wilms et al. (2000) are 
$7.6\times10^{-5}$, $8.71\times10^{-5}$ and $2.69\times10^{-5}$ for 
${\rm N}/{\rm H}$, ${\rm Ne}/{\rm H}$ and ${\rm Fe}/{\rm H}$ respectively. }
\tablenotetext{h}{Column density of once and twice ionized O towards Ark\,120, in units $\times10^{17}$\,cm$^{-2}$.}
\label{ism}
\end{deluxetable}

\clearpage

\begin{deluxetable}{lrcccc}
\tabletypesize{\small}
\tablecaption{Soft X-ray Emission Lines in the 2014 Ark\,120 Spectrum}
\tablewidth{0pt}
\tablehead{
\colhead{Line ID} & \colhead{E$_{\rm quasar}^{a}$} & \colhead{Flux$^{b}$} & \colhead{EW$^{c}$} 
& \colhead{$\sigma^{d}$} & \colhead{$\Delta \chi^{2}$$^{e}$}}
\startdata
N\,\textsc{vi} (i) & $428.4\pm2.3$ [28.941] & $9.6^{+4.5}_{-3.2}$ & $0.83^{+0.39}_{-0.27}$ & $4.9^{+1.9}_{-2.1}$ & 19.7 \\
N\,\textsc{vii} Ly-$\alpha$  & $501.0\pm0.4$ [24.747] & $1.8\pm0.5$ & $0.23\pm0.10$ & 
$<0.9$ & 13.1 \\
O\,\textsc{vii} (i)  & $568.2^{+1.7}_{-1.6}$ [21.820] & $18.8^{+3.3}_{-3.0}$ & $2.8\pm0.5$ & 
$6.8^{+1.5}_{-1.2}$ & 104.8 \\
O\,\textsc{vii} (f) & $561.0^{f}$ [22.100] & $2.6\pm2.2$ & $0.4\pm0.3$ & $1^f$ & --\\
O\,\textsc{viii} Ly-$\alpha$ & $654.1\pm0.5$ [18.955] & $2.1\pm0.7$ & $0.47\pm0.16$ & 
$<1.1$ & 27.2 \\
Fe\,\textsc{xviii} & $864.7\pm1.9$ [14.338] & $1.4\pm0.6$ & $0.60\pm0.25$ & $2.1^{+2.3}_{-1.5}$ & 15.4 \\
Ne\,\textsc{ix} (f) & $904.8\pm3.6$ [13.703] & $2.9\pm0.8$ & $1.4\pm0.4$ & 
$7.1^{+4.4}_{-2.7}$ & 36.6 \\
Mg\,\textsc{xi} & $1342\pm10$ [9.239] & $1.6^{+0.9}_{-0.6}$ & $1.9^{+1.1}_{-0.7}$ & 
$<24$ & 16.7 \\
Mg\,\textsc{xi}$^{g}$ (f) & $1333.5^{+1.3}_{-1.5}$ [9.297] & $2.4\pm1.1$ & $2.2\pm1.0$ & $2.0^{+1.3}_{-0.8}$ & 14.8 \\
Mg\,\textsc{xii} & $1475^{+1}_{-3}$ [8.406] & $1.8\pm0.6$ & $2.6\pm0.9$ & 
$<3.6$ & 28.1 \\
\enddata
\tablenotetext{a}{Measured line energy in quasar rest frame, units
  eV.  The corresponding rest wavelength value in \AA\ is given
within brackets. Note the lines are measured with the RGS, unless otherwise stated.}
\tablenotetext{b}{Photon flux of line, units $\times10^{-5}$\,photons\,cm$^{-2}$\,s$^{-1}$}
\tablenotetext{c}{Equivalent width in quasar rest frame, units eV.}
\tablenotetext{d}{$1 \sigma$ velocity width, eV.}
\tablenotetext{e}{Improvement in $\Delta \chi^{2}$ (or $\Delta C$ for the HETG) upon adding line to model.}
\tablenotetext{f}{Indicates parameter is fixed.}
\tablenotetext{g}{Parameters measured from the Chandra HETG spectrum}
\label{emission-lines}
\end{deluxetable}

\clearpage

\begin{deluxetable}{ccccc}
\tabletypesize{\small}
\tablecaption{Fits to Velocity Profiles of Soft X-ray Emission Lines}
\tablewidth{0pt}
\tablehead{
\colhead{Line ID} & 
\colhead{$\sigma_{\rm v}$$^{a}$} & \colhead{FWHM$^{a}$} 
& \colhead{$v_{\rm out}$$^{b}$} & \colhead{Flux$^c$}} 
\startdata
Single Gaussians$^{d}$:-\\
\hline
N\,\textsc{vi} He-$\alpha$ & $3200\pm1300$ & $7500\pm3000$ & $-1500\pm1000$ & $9.6^{+4.5}_{-3.2}$ \\
N\,\textsc{vii} Ly-$\alpha$ & $<290$ & $<680$ & $-410\pm120$ & $1.8\pm0.5$ \\
O\,\textsc{vii} He-$\alpha$ & $4050\pm700$ & $9500\pm1600$ & $<730$ & $18.8^{+3.3}_{-3.0}$ \\
O\,\textsc{viii} Ly-$\alpha$ & $<470$ & $<1100$ & $-250\pm130$ & $2.1\pm0.7$ \\
Ne\,\textsc{ix} He-$\alpha$ & $3980\pm1100$ & $9350\pm2600$ & $<710$ & $2.9\pm0.8$ \\
Mg\,\textsc{xi} He-$\alpha^{g}$ (f) & $450^{+290}_{-180}$ & $1050^{+680}_{-420}$ & $-540^{+340}_{-290}$ & $2.4\pm1.1$ \\ 
Mg\,\textsc{xii} Ly-$\alpha$ & $<700$ & $<1650$ & $<300$ & $1.8\pm0.6$ \\
\hline
Three Gaussian fit$^{e}$:-\\
\hline
N\,\textsc{vi} triplet:-\\
(f) & -- & -- & $-1650\pm400$ & $<3.6$ \\
(i) & $1840\pm820$ & $4300\pm1900$ & $-1450\pm700$ & $6.3\pm3.0$ \\
(r) & -- & -- & $<2300$ & $<2.5$ \\
\hline
O\,\textsc{vii} triplet:-\\
(f) & -- & -- &  & $6.9\pm2.8$ \\
(i) & $2000^{+1200}_{-800}$ & $4600^{+2700}_{-2000}$ & $<850$ & $6.5\pm2.5$ \\
(r) & -- & -- &  & $6.4\pm2.3$ \\
\hline
Ne\,\textsc{ix} triplet:-\\
(f) & $2300^{+1800}_{-1100}$ & $5500^{+4200}_{-2600}$ & $<930$ & $3.0\pm1.1$ \\
(i) & -- & -- & -- & $<0.6$ \\
(r) & -- & -- & -- & $<0.5$ \\
\enddata
\tablenotetext{a}{Intrinsic $1\sigma$ and FWHM velocity widths of emission lines in km\,s$^{-1}$, after correcting (in 
quadrature) for instrumental spectral resolution.}
\tablenotetext{b}{Velocity shift of emission line in km\,s$^{-1}$. Negative 
values denote blue-shift. The velocity centroids of the N\,\textsc{vi} and O\,\textsc{vii} lines are measured with respect 
to the intercombination emission and Ne\,\textsc{ix} is with respect to the forbidden line.}
\tablenotetext{c}{Flux of the individual He-like line components, in units $\times10^{-5}$\,photons\,cm$^{-2}$\,s$^{-1}$.}
\tablenotetext{d}{Profile fitted with a single Gaussian component only.}
\tablenotetext{e}{He-like profiles fitted with a blend of three broad Gaussian lines, with a common velocity width 
between the three components. 
Note f, i and r denote forbidden, intercombination and resonance components respectively.}
\tablenotetext{g}{Profile measured from the Chandra HETG spectrum}
\label{line-widths}
\end{deluxetable}

\clearpage

\begin{deluxetable}{lccc}
\tabletypesize{\small}
\tablecaption{Soft X-ray emission zones modeled by \textsc{xstar}.}
\tablewidth{0pt}
\tablehead{
\colhead{Zone} & \colhead{$\log\xi^{a}$} & \colhead{$F_{\rm cov}$ ($N_{\rm H}=10^{21}$\,cm$^{-2}$)$^{b}$} &  \colhead{$F_{\rm cov}$ ($N_{\rm H}=10^{22}$\,cm$^{-2}$)$^{b}$}}
\startdata
Broad (O\,\textsc{vii}) & $0.5\pm0.1$ & $0.36\pm0.06$ & $0.08\pm0.02$ \\
Broad (Ne\,\textsc{ix}) & $1.5\pm0.3$ & $>0.64$ & $0.09\pm0.03$ \\
Narrow (O\,\textsc{viii}) & $2.3\pm0.4$ & $>0.62$ & $0.11\pm0.04$ \\
\enddata
\tablenotetext{a}{Ionization parameter. Units of $\xi$ are erg\,cm\,s$^{-1}$ in log units.}
\tablenotetext{b}{Covering fractions for column densities of $N_{\rm H}=10^{21}$\,cm$^{-2}$ and $10^{22}$\,cm$^{-2}$ respectively.}
\label{xstar}
\end{deluxetable}

\end{document}